## Hydrogen Bonding in Water Under Extreme Confinement

*Xintong Xu[1], Matthias Kuehne[2,3], Harrison A. Walker[4,5], De-Liang Bao[4], Xin Jin[4,5], Yu-Ming Tu[2], Cody L. Ritt[2], Joel Martis[1], Juan Carlos Idrobo[6,7], Sokrates T. Pantelides[4,8], Michael S. Strano[2], Jordan A. Hachtel[*9], Arun Majumdar[*1,10]*

[1]Department of Mechanical Engineering, Stanford University, Stanford, CA, USA
[2]Department of Chemical Engineering, Massachusetts Institute of Technology, MA, USA
[3]Department of Physics, Brown University, Providence, RI, USA
[4]Department of Physics and Astronomy, Vanderbilt University, Nashville, TN, USA
[5]Interdisciplinary Materials Science Program, Vanderbilt University, Nashville, TN, USA
[6]Materials Science & Engineering, University of Washington, Seattle, WA, USA
[7]Physical and Computational Sciences Directorate, Pacific Northwest National Laboratory, Richland, WA, USA
[8]Department of Electrical and Computer Engineering, Vanderbilt University, Nashville, TN, USA
[9]Center for Nanophase Materials Sciences, Oak Ridge National Laboratory, Oak Ridge, TN, USA
[10]Department of Photon Science, SLAC National Laboratory, Menlo Park, CA, USA

[*]Correspondence to: arunava@stanford.edu
[*]Correspondence to: hachtelja@ornl.gov

**Fluids under extreme confinement exhibit unique structures and intermolecular bonding, distinct from their bulk analogs, driving innovative applications at the water-energy nexus[1–4]. Probing confined water experimentally at the length scale of intermolecular and surface forces has, however, remained a challenge. Here, we report direct molecular-level observations of hydrogen bonding in water confined inside individual carbon nanotubes, enabled by in-situ vibrational electron energy-loss spectroscopy with nanoscale resolution. Hydrogen bonding is probed via the intramolecular O–H stretching frequency, which serves as a sensitive spectral signature of the local intermolecular bonding environment. Water in larger carbon nanotubes exhibit the bonded O-H vibrations of bulk water, but at smaller diameters, the frequency blueshifts to near the free O-H stretch found in water vapor and hydrophobic surfaces, indicating a highly dispersed, non-H-bonded environment. Theoretical analysis based on quantum vibrational oscillators indicates that enhanced damping rates, corresponding to rapid hydrogen-bond fluctuations, leads the bimodal spectral peaks to merge into a single broad feature, matching the experimental observation. Furthermore, cryogenic experiments provide insights into complex structural phase transitions of confined water. This research reveals the quantum and dynamic nature of hydrogen bonds under confinement and the potential impact of unveiling molecular-level structure and bonding in confined fluids.**

## Introduction

Exotic behaviors in fluids emerge under extreme confinement in nanoscale channels, cavities, and porous media, where the length scales match the fundamental range of intermolecular forces. Carbon nanotubes (CNTs) with atomically smooth inner surfaces and diameters in the range of 1-2 nm, resemble biological water channels, and give rise to significant enhancement of water transport, orders of magnitude faster than that in bulk[1–4]. The anomalous behaviors of confined water arise from changes in molecular structure, depicted by the hydrogen-bond (H-bond) network - a subtle interplay of localized bonding and long-range interactions, which gives rise to diverse phases of water layers at interfaces[5–7]. The bonding is particularly sensitive to confinement due to the light mass of hydrogen, whose motion is inherently quantum mechanical and dynamically fluctuating[8,9]. Deeper understanding of the complex structure and bonding of water in CNTs and the interactions of confined fluids with their bounding surfaces could facilitate new technologies at the water-energy nexus[10–12].

Directly probing the H-bond network of confined water at the molecular-level has remained challenging. Experimental examinations of water in CNTs have mostly been conducted with macroscopic optical and beam-line techniques on ensembles of CNTs with non-uniform diameter[13–16]. These techniques provide robust statistics, and with careful experimental design, can limit the influence of CNT polydispersity[16]; however, they cannot assign specific behaviors to individual CNTs. Even in optical experiments that leverage sparsity to isolate CNTs from one another[17], the limited cross-section of a single tube prevents direct analysis of water within. Attempts to improve spatial resolution have been made with transmission electron microscopy (TEM), however, the low scattering cross-section of hydrogen and the dynamic nature of water make direct imaging exceptionally challenging[18].

Molecular dynamics (MD) simulations have been the main approach to investigate the behavior of confined water at the molecular level. They have predicted a range of confinement-dependent structural phases of water in CNTs, ranging from bulk-like water in large-diameter CNTs ($d$>2 nm) to complex structures in medium-diameter CNTs ($d$~1-2 nm) to one-dimensional chains in small-diameter CNTs ($d$<1 nm)[13,15,19–28]. For medium-diameter CNTs, the predicted networks vary across the literature, spanning highly ordered 'ice tubes' to disordered arrangements. Simulated vibrational spectra[19,24] often suggest coexisting vibrational modes at ~420 meV ('bonded' O–H stretch, as in bulk water) and ~460 meV ('free' O–H stretch, as in water vapor[29] and hydrophobic interfaces[5]). However, they lack direct validation by experiments capable of probing vibrational properties of water within individual CNTs.

A promising option for probing confined water at the molecular scale is vibrational electron energy-loss spectroscopy (vEELS) in a monochromated scanning transmission electron microscope (STEM). The technique can achieve energy resolution as low as ~3 meV (25 cm$^{-1}$), while retaining sub-nanometer spatial precision[30]. Moreover, vEELS possesses a high sensitivity down to a few molecules enabling direct measurement of molecular vibrations[31–36], including the O–H stretching mode of water[37]. Because the frequency of the intramolecular O–H stretch is strongly mediated by intermolecular H···O interactions[6], it serves as a sensitive probe of the local hydrogen-bonding environment.

**Visualizing Water in Individual CNTs with vEELS**

A schematic of the vEELS experiments is shown in Fig. 1a. The CNTs are directly grown on $SiN_x$ TEM grids, with a periodic array of 200 nm holes as the electron-beam window, by chemical vapor deposition (CVD) to avoid contamination during transfer. After grown tubes are cut open by focused ion beam (FIB), they are placed in a humidifier to allow water to infiltrate them (shown schematically in Supplementary Fig. S1). The CNTs are sparse, millimeters long, and cross the membrane in parallel[38] as shown in Fig. 1b, making them easy to identify and isolate as well as leaving the residual contaminants on the membrane far away from the measurement window. Reference images are conducted in STEM mode with a high-angle annular dark-field (HAADF) detector, as shown in Fig. 1c, which allows us to precisely control the beam for vEELS acquisitions. To minimize the dose, we primarily acquire point spectra through the aloof mode, where the beam is placed closed to the sample (~30 nm), without directly intersecting it, and the vibrations are probed through the evanescent interactions[39]. We also acquire spectrum images (SI), where vEEL spectra are acquired at each probe position in a 2D grid to create a full three dimensional (two spatial, one spectral) hyperspectral dataset, which enables localization of the vibrations to be directly measured. Here, the dose is minimized by scanning over a large area with low dwell times, ensuring no part of the sample is continually irradiated with the electron beam.

Figure 1d shows aloof vEEL point spectra recorded from two different nominally water-filled CNTs compared against vEELS recorded from a bulk liquid cell. One CNT (bottom spectrum) appears empty, as no peak is observed in the O-H stretching regime. However, the other CNT (top spectrum) shows a dominant peak centered at 420 meV, which is consistent in both linewidth and frequency to the bulk-water O-H stretch frequency obtained from the liquid cell. The match verifies the existence of water in the CNT and demonstrates the capacity of vEELS to directly interrogate water vibrations in a confined CNT environment.

SI are acquired from the empty and filled tubes from Fig. 1d. The simultaneously acquired HAADF images are shown in Fig. 1e (filled) and Fig. 1f (empty) exhibiting no clear differences. By taking a SI slice of the O-H stretch region (integrating the spectrum in each SI pixel between 400 and 500 meV), we can see in Fig. 1g that the filled tube lights up, while Fig. 1h shows that the empty tube has barely any variation from the background.

**The Effect of Extreme Confinement on H-Bonding**

To explore the effect of confinement on H-bonding, we probed the frequency of O-H stretching modes in water-filled CNTs with different diameters. Figure 2a compares the room temperature (300 K) vibrational spectra of water confined in CNTs with inner diameters (*d*) of 2.3 and 1.4 nm. TEM images of the tubes confirming the diameters are shown in the Extended Data Figure 1. A significant difference is observed: The 2.3 nm-diameter CNT spectrum matches with that of bulk water, *i.e.*, a single vibrational peak at 420 meV (3400 $cm^{-1}$), while in the more confined system (1.4 nm diameter CNT), the vibrational peak blue-shifts to ~455 meV (~3700 $cm^{-1}$). This is the free O-H stretch frequency observed in water vapor or at hydrophobic water-air and water-oil surfaces/interfaces where the symmetry in the H-bond network of the system is broken[5,29]. The fact that the vibrational response is dominated by the free O-H stretch mode suggests that the water molecules are weakly or non-H-bonded, demonstrating the emergence of an unusual structural phase of water under extreme confinement characterized by significant disruption of the H-bond network.

To estimate the vibrational contributions qualitatively, we performed a two-Gaussian deconvolution of the spectra, assigning components near 420 meV and 455 meV to bonded and free O–H modes, respectively. In the 2.3 nm-diameter CNT, ~90% of the spectral weight resides in the bonded mode, whereas in the 1.4 nm-diameter CNT, this population is reversed, with ~86% of the signal arising from the free O–H stretch. The presence of free O-H aligns with predictions in simulations attributing such signatures to steric constraints of hydrophobic CNT that produce a partial breakdown of the H-bond network[27,28]. On the other hand, the observed spectrum appears as a single broad peak with predominant population of free O-H rather than a bimodal distribution, indicating a dynamically fluctuating environment rather than a static coexistence of distinct free and bonded modes. We note that the vEELS energy-resolution (~8.5 meV) is demonstrably sufficient to distinguish these features (Extended Data Figure 2).

To understand the origin of our observations, we developed a theoretical model based on quantum Morse oscillators[40–42]. Figure 2c illustrates the computed potential energy curves and quantized vibrational levels for the free and bonded O–H oscillators. The energy loss functions (ELFs) are constructed based on this mode and shown in Fig. 2b. The free O–H intramolecular stretch is modeled by a conventional Morse potential with a deep, narrow well. In contrast, the bonded mode is represented by a Morse potential perturbed by a neighboring oxygen atom (H···O intermolecular interaction), represented by the superposition of a second Morse term. This results in a shallow, broadened double-well potential near the equilibrium bond length. By numerically solving the one-dimensional Schrödinger equation, we obtained the corresponding vibrational eigenstates, which become more delocalized and exhibit reduced level spacing, leading to a redshifted transition frequency. The dielectric response of the system is then constructed from two vibrational resonances with weighted populations, which can be used to create a model for ELFs shown in Fig. 2b[43]. Full expressions and parameters used in the calculations are provided in Methods.

The calculated ELFs utilize the experimentally estimated population ratios and varying damping rates to elucidate the effect of hydrogen-bond dynamics. The top panel in Fig. 2b shows the ELF for a model with 90% bonded mode, matching the experimental spectrum from the 2.3 nm CNT. The bottom panel shows the ELF with 86% free OH, corresponding to the 1.4 nm case. The damping factor ($\Gamma$) is the dimensionless quantity related to the inverse of the characteristic vibrational lifetime. As $\Gamma$ increases, the bimodal spectrum merges into a single broad one. This evolution captures the influence of rapid hydrogen-bond fluctuations, which mix vibrational modes and thereby distribute the energy between water molecules. Ultrafast vibrational sum frequency generation (vSFG) spectroscopy of interfacial water[44], though not yet extended to confined systems, showed that the free OH modes can transfer their vibrational energy to the H-bonded OH modes before intramolecular decay. Under confinement, we suggest that hydrogen bonds between water molecules are not merely broken but transient, disrupted by continuous intermolecular reorganization with a large fraction of free-OH, creating a dynamically disordered, non-H-bonded environment. The dynamic nature of the water-CNT system is illustrated schematically in Fig. 2d.

What is clear in Fig. 2a is that the widths of the vEEL spectra of water inside the 2.3nm- and 1.4nm-diameter CNTs are roughly the same. However, resolving the detailed pathways of energy exchange and relaxation for water under confinement will require accurate quantum molecular dynamic simulations[45].

It is noteworthy that prior MD simulations, along with new ones reported in Supplementary Figure 5, do not accurately replicate the experimental observations for the 1.4- and 2.3-nm-diameter CNTs. Instead, they predict a more even ratio of bonded and free O-H in the CNTs, with distinct peaks for both the vibrational modes more akin to the $\Gamma$=1 or $\Gamma$=2 cases in Fig. 2b as opposed to the $\Gamma$=4 case that matches the experiment more closely. For the 1.4-nm-diameter CNT, MD simulations did show that at significantly elevated temperatures (~500 K) the bonded O-H peak merges with the free O-H peak to form a broad singular distribution, similar to the observed distribution in vEELS. However, since the spectra can be obtained in the aloof configuration (where the beam does not directly intersect with the sample) it is highly unlikely that beam-induced heating could generate a 200 K increase. This disparity highlights the complexity of the water-tube interaction and suggests future theoretical research is needed to explain the observed behavior.

**The Temperature-Dependence and Spatial Heterogeneity in Confined Fluids**

For confined water, subtle variations in the equilibrium between molecule-molecule and molecule-channel interactions can fundamentally change the phase diagram, since they are of comparable strength. This competition opens many potential variations in the H-bond network and new phases of water and ice[46,47]. Here, we show initial results by performing in situ vEELS measurements while lowering the temperature from 300 K to 100 K, followed by reheating back to 300 K to complete a thermal cycle.

In the 2.3 nm-diameter CNT, which exhibits a single bonded O–H stretch peak at 420 meV at room temperature, cooling to 100 K induces a secondary feature near 400 meV (Fig. 3a). This change disappears upon warming back to 300 K (Extended data Fig. 3), confirming the reversibility. Figure 3b shows the spectrum image slice of the O-H intensity (between 370-500 meV) of the 2.3- nm-diameter CNT at 100 K, with the spatially resolved line profile along the CNT length shown in Fig. 3c. The newly emerged low-frequency mode aligns with the characteristic vibration of tetrahedrally coordinated ice Ih at 400 meV[13], while the 420 meV peak resembles that of bulk amorphous ice[48]. The coexistence of both features is consistent with a mixture of ice phases, as previously predicted for layered nanotube-ice systems[47]. The results are also in line with theoretical predictions of a non-equilibrium phase diagram for supercooled water, in which two distinct amorphous ice phase and a liquid phase can coexist[49]. The line profile reveals a relatively largely homogeneous spectral evolution along the nanotube length, with a slight blueshift indicative of locally weaker hydrogen bonding, comparable to the blueshift reported for "nanotube ice" in earlier neutron scattering work[13]. These results suggest a partial phase transition driven by thermally modulated reorganization of the H-bond network into a complex, inhomogeneous ice-like state.

The 1.4-nm-diameter CNT, which exhibits a dominant free OH mode at 455 meV at 300K (Fig. 3d, top), shows spatially heterogeneous behavior upon cooling. At 100 K, some regions retain the high-frequency peak, while others develop a bimodal spectrum with an additional component near 420 meV (Fig. 3d, middle R1 and bottom R2). The spectrum image slice of the O-H intensity is shown in Fig. 3e. Gaussian deconvolution confirms the emergence of a bonded O–H signature in select regions. This partial recovery of hydrogen bonding under cooling reflects the metastable and spatially disordered nature of confined water, as well as the possibility of suppression of rapid H-bond fluctuations at room temperature. These findings build on the picture established in Figure 2: at cryogenic temperatures, damping is reduced and dynamic disorder partially arrested, allowing incipient static ordering to emerge. Nevertheless, the changes between 300 K and 100 K are overall subtle, with the high-frequency peak largely retained. The absence of a peak at 400 meV suggests that major rearrangement of H-bond network in the confined system is impeded with most free-OH preserved. Similar effects have been observed in nano-droplets, mesoporous membranes, and the Raman spectra of filled CNTs, where reformation of the H-bond network is suppressed when lowering temperature for small droplet diameters and pore sizes[17,50,51].

Finally, we present preliminary results in the 0.8 nm confinement regime. Purified bundles of 0.8 nm CNTs are filled at room temperature, frozen in liquid nitrogen and cryo-transferred into the electron microscope at 100 K. Figure 3g shows a dark field image of a bundle of 0.8 nm CNTs, overlaid with masks of where the free O-H signature at 455 meV is observed (blue) and where it is absent (gray), where the masked areas are determined by the intensity of the free O-H stretch in a per-spectrum two peak fit at the bonded and free O-H stretch frequencies shown in Fig. 3h. The bimodal spectrum with free OH signature shows up besides the dominant bonded OH (blue), possibly suggesting the existence of the single-file water long predicted[27]. The spatially heterogeneous vibrational responses are further consistent with the above described dynamic, complex nature of H-bond network in ultrasmall systems. Advantages and limitations of this sample preparation are discussed in Methods.

Leveraging the high energy and spatial resolution of vibrational EELS, along with cryogenic capabilities, we directly probe molecular-scale vibrational signatures. Our results reveal how the structure and bonding of water are fundamentally altered under nanoscale confinement inside carbon nanotubes of different diameters. The majority of intermolecular hydrogen bonds are disrupted, potentially by rapid fluctuations and energy exchange, broadening the vibrational spectra of water. The confinement size and temperature-dependency synergistically modulate the delicate competition between molecule–molecule and molecule–wall interactions. Combined with quantum theoretical analysis, our findings uncover the transient, dynamically fluctuating nature of hydrogen bonding under confinement, offering a powerful framework for understanding confined fluids in complex nanoscopic environments. In nanofluidics, the state of the H-bond network is an essential element in the behavior of solutes and the flow of water, impacting applications in desalination, biotechnology, and drug delivery. The phase boundaries of such networks are predicted to depend on confinement variables such as pore diameter and wall potential, and not just isobaric temperature[11]. The workflow exemplified here offers a new paradigm in the analysis of confined fluids in complex systems, and can be extended from water in 1D channels (such as CNTs) to fluids or ions confined in 2D channels such as van der Waals gaps or even 0D pores such as in zeolites[52,53], and impact technologies such as batteries, sensors, catalysis, and hydrogen storage.

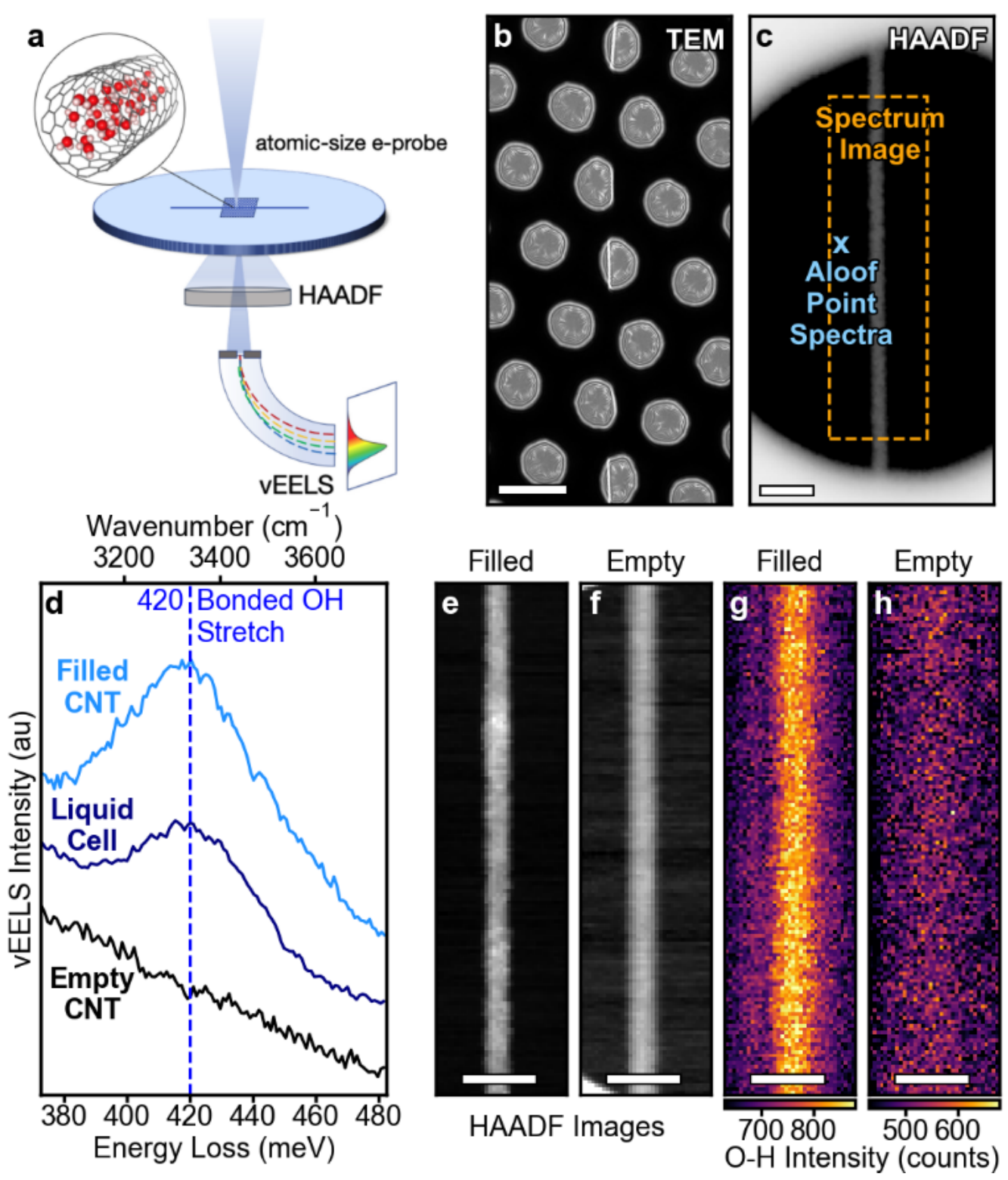

**Fig. 1. Visualization of water confined in a single carbon nanotube.** (a) Schematic of the vibrational electron energy-loss spectroscopy (vEELS) experimental setup. (b) Low-magnification TEM image of nanotubes on 200 nm diameter holey SiNx grid. Scale bar is 250 nm. (c) HAADF image of a CNT. Annotations denote probe position for aloof point spectra, and the region for a spectrum image. Scale bar – 50 nm. (d) Aloof vEEL spectra of a filled CNT (top), a bulk-water liquid cell (middle) and an empty CNT (bottom). (e,f) HAADF images of filled (e) and empty (f) CNTs. (g,h) Corresponding spectrum image slices integrating the intensity of the O-H stretch vibrations (between 400 and 500 meV) for the filled and empty CNTs. Scale bars – 20 nm. The filled CNT was measured to have an inner diameter of 2.3 nm.

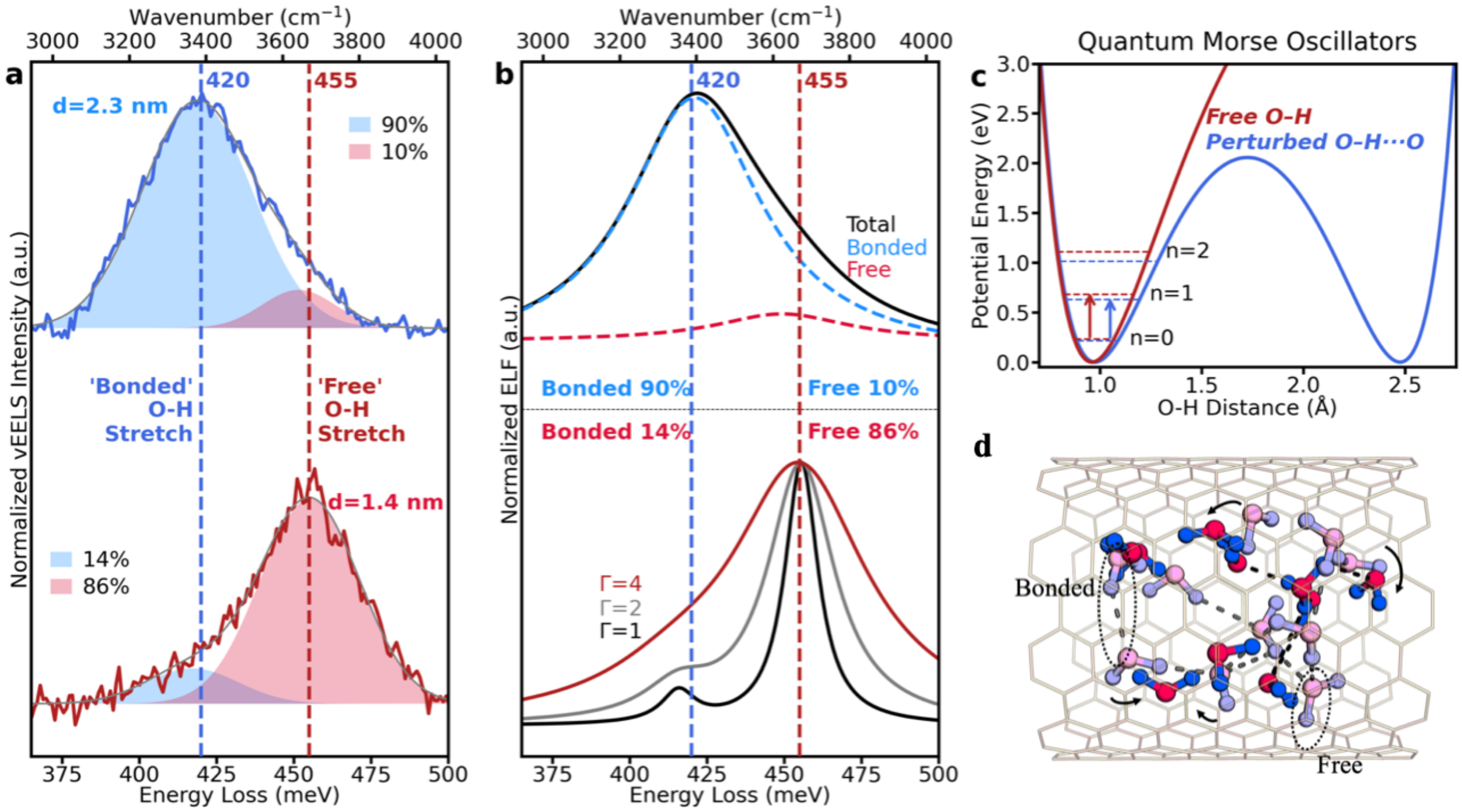

**Fig 2. Vibrational properties and hydrogen bonding of water confined in medium diameter CNTs at room temperature.** (a) vEEL spectra from tubes with different inner diameters: 2.3 nm – top (same as Fig. 1) and 1.4 nm – bottom. The larger tube exhibits primarily the 'bonded' O-H stretch, i.e., the O-H···O stretch at 420 meV (3400 $cm^{-1}$) while the smaller tube exhibits primarily the 'free 'O-H stretch frequency at 455 meV (3670 $cm^{-1}$). The spectra are deconvolved using two Gaussian functions to qualitatively estimate the relative populations of bonded and free O–H groups. (b) Calculated vibrational energy loss functions (ELFs) based on quantum Morse oscillators. The top panel shows the total ELF for a 10% free O–H case, along with decomposed contributions from bonded and free oscillators. The bottom panel shows the ELF for an 86% free O–H case with varying damping rates ($\Gamma$ = 1, 2, 4), illustrating how increased damping rate (inverse of characteristic lifetime) merges the bimodal peaks into a single broad mode. ($\Gamma$ = 4 used for top panel.) (c) Morse potentials as a function of O–H bond length for free and perturbed (hydrogen-bonded) O–H···O configurations. The perturbed potential is modeled as a double-well Morse oscillator, reflecting intermolecular hydrogen bonding. Quantized vibrational energy levels and fundamental ($\Delta n$ = 1) transitions are overlaid. (d) Schematic of water confined in a CNT (side view), showing two overlapping ensembles of water molecules representing different timeframes, illustrating rapid hydrogen-bond rearrangement. O/H atoms are colored red/blue and pink/purple for two ensembles, respectively, with dashed lines indicating H bonds. Free and bonded O–H groups are marked.

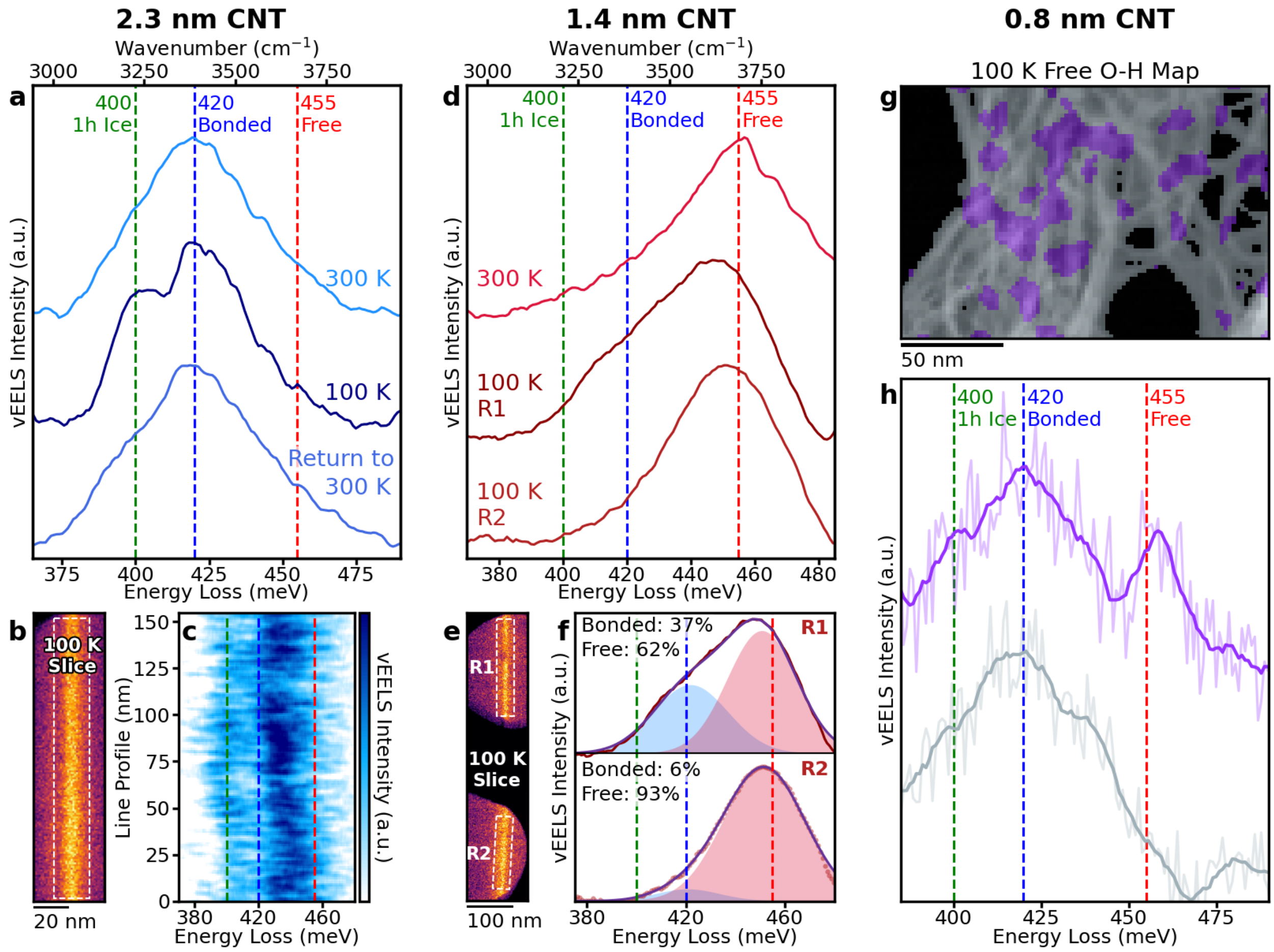

**Fig 3. Temperature-dependent vibrational responses of water-filled CNTs.** (a) vEEL spectra for a 2.3 nm diameter CNT at 300 K and 100 K, showing a temperature-induced change of the main vibrational responses. Vertical dashed lines mark reference positions for tetrahedrally bonded ice (~400 meV), hydrogen-bonded (~420 meV), and free (~455 meV) O–H stretches. (b) Spectrally integrated (between 370-500 meV) image at 100K of the water-filled 2.3nm-diameter CNT. (c) Vibrational spectra of water inside the 2.3 nm diameter CNT at 100K and along the CNT length. (d) vEEL spectra from a 1.4 nm diameter CNT at 300 K (top) and at 100 K in two spatial regions (middle R1; bottom, R2), revealing heterogeneity at low temperature. Gaussian deconvolution quantifies the relative populations of bonded and free O–H modes. (e) Spectrally integrated (between 370-500 meV) image at 100 K of the water-filled 1.4 nm-diameter CNT in two different regions R1 and R2. (f) The averaged spectra at R1 and R2 with Gaussian decovolution show different ratios of bonded and free OH vibrations. (g) Free O–H EELS region map of a 0.8 nm CNT bundle at 100 K, highlighting areas with (blue) and without (grey) free O–H vibrational signatures. (h) Corresponding vibrational spectra from the highlighted regions in (g) with blue showing free O-H stretch at about 455 meV and grey showing no free O-H stretch but exhibiting bonded O-H stretch at 420 meV.

**Supplementary Information**

Supplementary Information is available for this paper.

## Methods

*Sample Preparation:* K-kit $SiN_x$ liquid cells (Silicon-based Micro Channel Device) were purchased from Electron Microscopy Sciences (EMS) with membrane thickness of 30 nm and gap height of 100 nm. The liquid cells were opened, filled with DI water, and sealed with TorrSeal following the instructions provided by EMS. Ultralong CNTs were grown using a low-gas-flow variant of chemical vapor deposition (CVD)[1], as previously described[2,3], with growth performed directly on holey $SiN_x$ TEM grids (500 nm or 200 nm pores; TED Pella, 21583-10 or 21586-10, respectively), resulting in a sparse array of gas flow-aligned ultralong CNTs[4]. For this work, $CH_4$ was utilized as the carbon precursor, and iron nanoparticles embedded in 25 Series APT carbon nanotubes (Nano-C dispersion, prepared following reported procedures[5]) served as catalysts. Catalyst dispersions were drop-casted onto the upstream side of the growth substrates. During CVD, substrates were positioned in a tube furnace such that the catalyst layer faced upstream of the SiNx membranes, allowing gas-flow–aligned ultralong CNTs to grow across the grid and suspend over the open windows[4].

After CVD synthesis, the CNTs were then subjected to a previously introduced Focused Ion Beam (FIB) cutting procedure to induce CNT opening[6,7]. Briefly, the focused $Ga^+$ beam of a dual beam FIB/scanning electron microscope (FEI Helios Nanolab 600) is used to cut the ultralong CNTs between the catalyst and TEM window and once beyond the same window. We use the following

FIB patterning parameters: 9 pA, 30 kV, 100 passes, 5 μs dwell time, and 4 nm pitch. Any exposure of the membrane-supported CNT segments in between the two cut locations to electron or ion beams was carefully avoided. The CNTs were then transported and exposed to above 99% RH conditions in a custom-built humidity chamber (Supplementary Fig. S1a) using an ultrasonic humidifier as a standardized water filling process. We subsequently applied TorrSeal (Agilent Technologies) to the two FIB-cut regions of the sample (Supplementary Fig. S1b), again avoiding the region between, as a CNT sealing method. Distinctive, temperature dependent Raman RBM trajectories were observed for CNTs before and after the water filling process, as described elsewhere[2,7].

Single-walled (6,5), double-walled and multi-walled CNTs (DWCNTs and MWCNTs) were purchased commercially from Millipore Sigma. The (6,5) tubes were product number 773735, with approximately 41% of those tubes being (6,5) chirality with diameter of 0.8 nm. The DWCNTs were product number 637351-250MG which had nominal dimensions of outer diameter × inner diameter × length of 5 nm × 1.3-2.0 nm × 50 μm. The MWCNTs were product number 698849-1 with nominal dimensions of outer diameter × length of 6-13 nm × 2.5-20 μm. Samples or validation tests were prepared by sonication and drop casting onto lacey carbon grids. The grids loaded with (6,5) tubes were blotted by filter papers, then frozen in liquid nitrogen for cryogenic experiments.

Large bundles of 0.8nm-diameter single-walled CNTs with selected (6,5) chirality are opened and filled via sonication (>~40%), then drop casted onto a lacey carbon TEM grid and cryo-transferred into TEM. There are several key difficulties with this strategy. Firstly, the nature of the sample preparation prevents from isolating individual CNTs. Additionally, room temperature experiments are challenging as lacking ways to directly seal the CNTs after filling (and hence at RT the water escapes in the vacuum of the microscope). Lastly, the residual water outside of CNTs is trivial since they are hydrophobic, however, some may still be frozen and contributed to bonded O-H signal. However, there are also some key benefits of this strategy. These samples represent the highest level of confinement possible in the water-filled CNT system suggested the possibility of the most pronounced free O-H stretch signature. Lastly, the signature of spatially varied free O-H stretch inside the CNT offer us a marker of where the confined water is present in the system.

*Transmission Electron Microscopy:* TEM images were acquired on an FEI Titan 80-300 environmental transmission electron microscope operated at 80 kV, with an aberration corrector for the objective lens tuned before each session. Inner diameters are measured from the half-maximum of the inner most interference fringe in the TEM image, the error in this measurement is estimated to be approximately half of the interference fringe width (0.34 nm) indicating an effective ±0.1 nm uncertainty. The electron dose rate during HRTEM image acquisitions was measured at ~300 $e^{-}/Å^{2}/s$. TEM images were analyzed using GATAN DigitalMicrograph and ImageJ software.

*Vibrational Electron Energy Loss Spectroscopy (Experimental Details):* The vibrational EELS in this manuscript is acquired with the Nion high energy-resolution monochromated EELS-STEM (HERMES) located at Oak Ridge National Laboratory. Data were acquired over eleven different samples, operating at a 60 kV accelerating voltage. The 1.4 nm-diameter CNT and the 2.3 nm-diameter CNT are acquired from the same sample during the same session to insure minimum experimental variance. All spectra acquired with a convergence angle of 30 mrad and a collection angle of 25 mrad. Experiments were attempted at 30 kV to improve energy resolution, but it was found that the widths of the spectral features are sufficiently broad whereby there is little benefit to improving beyond ~8 meV. Furthermore, the additional charging of the lower kV beam on the insulating substrate made experiments more challenging.

Energy resolution is defined by the full-width at half-maximum of the elastic scattering or zero-loss peak (ZLP). The point spectra in Fig. 1d had energy resolutions of 8.5 meV for both CNTs, and for the bulk liquid cell. The spectrum-images had energy resolutions of 9.5 meV for the filled CNT dataset, and 8.5 meV for the empty CNT dataset. The beam current for these experiments monochromation is approximately 4 pA, and while acquiring reference images to prepare spectrum images or point spectra, dwell times are kept extremely low to minimize sample irradiation.

For the spectrum images in Main Text Figure 1, we compare 100 meV wide slices from two different spectrum-images to demonstrate the strength of the O-H vibrations in the filled tubes. For the filled CNT in Fig. 1a 181 x 75 pixel spectrum-image was acquired with a 20 ms dwell time, for the empty tube a 151 x 78 pixel spectrum image was acquired with a 10 ms dwell time. In order to normalize, we use the maximum intensity of the zero-loss peak (ZLP) in the pixel furthest away from the CNT, and find that the empty tube dataset had a total current a factor of 3.12 times less than the filled tube current. For the 'counts' reported in Main Text Fig. 1 for the empty tube we have multiplied the total counts in the empty tube slice by this factor of 3.12 to make it directly comparable to the filled tube slice, the filled tube slice intensity is as acquired. The spectra are not background subtracted to make sure the difficulties described in the following section do not influence the slices. Additionally, isotropic stage drift causes all CNTs to appear diagonal during acquisition, which is corrected for the displayed spectrum-image slices.

The vibrational spectra in this system are complex in nature and possesses significant peaks both at lower energy and higher energy than the O-H stretch frequency range (400-460 meV) that are attributed to combination bands (Extended Data Figure 3) or excitons (Extended Data Figure 4). To isolate the O-H stretch and look exclusively at the 350-500 meV range, requires a complex background normalization/subtraction process. In core-loss EELS, background subtraction is conventionally done with a power law; however, it has been demonstrated that in the low-loss region the ZLP tail is often better fit with more complex functions[8,9]. Here, we use the two-region, third-order exponential fit described in Ref. 8, which possesses the form $I(E) = e^{ax^3+bx^2+cx+d}$. The fitting regions are set before and after the O-H stretch in order to isolate it from the peaks on either side. An example of the fitting, compared to a standard power-law background subtraction, is shown in Supplementary Fig. S2. While this methodology produces a non-physical background, it is only used as a means here of isolating the O-H stretch peak and is the best methodology we have found at capturing the line-shape and frequency of the O-H stretch peak.

Cryogenic experiments were conducted with a Gatan ELSA cryogenic cooling holder. All experiments are conducted in ultrahigh vacuum which minimizes ice buildup during cryo-experiments and prevents bulk water from staying adsorbed to the outside of the CNT (which could potentially cause anomalous measurements of the bonded O-H stretch). The lack of bulk-water-like peaks in the empty tubes shown in Fig. 1 and Extended Data Figure 3 and the lack of ice-like peaks in the 1.4-nm-diameter CNT cryo experiment shown in Fig. 3a serves as proof that ice is not building up significantly during the experiments.

*Vibrational Electron Energy Loss Spectroscopy (Peaks Beyond the O-H Stretch):* In the main text, we focus exclusively on the O-H stretch region between 370-500 meV. The lower energy peaks close to and below the O-H stretch peak are assigned as combination bands, similar to other IR absorption spectroscopy techniques, as well as other vEELS experiments on ice have reported peaks in these regions[10,11]. We have not rigorously determined the origin of these bands, but believe the response is largely consistent with combination bands/overtones.

In Extended Data Figure 3, we show the full range of the three spectra from the main text Figure 1 from a filled CNT (Top), an empty CNT (Middle), and the bulk liquid cell (Bottom). In the full range, the empty tube shows no features, and only a broad sloping continuum, consistent with that found in any conductive material, while the filled CNT shows many tightly packed peaks across the entire infrared. In the bulk liquid cell, the only peaks observed are the bonded O-H stretch, as expected, and the C-H stretch originating from contamination on the surface of the liquid cell, which could not be baked to suppress contamination or probed in aloof mode to avoid contamination (this is consistent with previous work on water in liquid cells[12]).

We also note that the empty CNT and the filled CNT are from the same sample. Thus, while polydispersity could potentially cause differences, the environmental conditions (such as contamination) should be identical. This serves as a strong piece of evidence that the peaks assigned as O-H stretch in the main text truly correspond to water, as it is difficult to envision a physical scenario that could result in such a dramatic difference between the vibrational spectra on nominally similar CNTs.

Beyond combination bands, additional peaks can come from excitons in semiconducting CNTs[13,14]. Extended Data Figure 4 shows the vibrational and valence regions for a filled CNT, compared to unfilled commercial double-walled (DW) and multi-walled (MW) CNTs. The filled CNT and the DWCNTs exhibit strong peaks at higher energies in the valence region, while the MWCNTs do not. Some of the DWCNTs exhibit weak peaks in the O-H stretch region, making differentiation between vibrations and excitons critical.

The peaks are differentiable in several ways. First is how common the lower energy vibrational modes attributed to the O-H stretch and combination bands are observed in the filled CNTs compared to the empty CNTs. The peak frequencies are tabulated for all CNTs in Extended Data Figure 4b and 4c. In the filled CNTs, 48% of the observed peaks are below 500 meV, while only 28% are below 500 meV in the unfilled CNTs. Furthermore, while some unfilled DWCNTs do exhibit peaks as low as 400 meV, none reach the 200-350 meV region, where we commonly see sharp peaks in the filled CNT samples, and none show peaks at the free O-H stretch which is the most observed frequency in the filled CNTs.

Secondly, there is a well-defined relationship between the CNT diameter and the excitonic frequencies, that is shown to apply to single wall and double wall CNTs[13,15,16]. An empirical formula for this relationship in air-suspended CNTs (Eq. 1), has been adapted from Ref. 15 and is plotted in Extended Data Figure 4d.

$$E_p(d) = 2\hbar v_f(p) \times \frac{2p}{3d} + \beta \times \frac{4p^2}{9d^2} + \eta(p) \times \frac{4p^2}{9d^2} \times \cos(3\theta) \qquad (1)$$

Here, $p$ refers to the order of the excitonic transition (*i.e.* $p$=1 corresponds to the $S_{11}$ mode and $p$=2 corresponds to the $S_{22}$ mode), we only examine these two cases as they are the only types of excitations that can realistically reach the energies probed in these experiments for the diameters considered here. The terms $v_f(p)$, $\eta(p)$, and $\beta$ represent the Fermi velocity, an anisotropy prefactor, and a fitting parameter and have been determined experimentally from extensive experiments highlighted in Ref. 15. Here, $\theta$ corresponds to the chiral angle of the CNT and varies between 0° and 60°. We chose 30° as a midway point to be representative of the chiral range, but different $\theta$ only result in a maximum difference of 10 meV from the 30° value in the diameters considered here.

In the filled CNTs (Extended Data Figure 4e-4g), the relative frequencies of the peaks do not follow the predicted relationship, while in the unfilled CNTs (Extended Data Figure 4h-4j) they do. Moreover, the diameters in the filled CNTs are also not consistent with the excitonic peaks. The DWCNT in Fig. Extended Data Figure 4g is the DWCNT from the main text with an inner diameter of 1.4 nm and an outer diameter of 2.1 nm. However, the lowest energy peak is at 353 meV which would require a diameter of 2.98 nm. Conversely, for the unfilled CNTs, the predicted diameters are all in line with the expected inner diameter range in the specifications from Millipore-Sigma which is 1.3-2.0 nm.

We also note that the excitons in the unfilled DWCNTs all correspond to the inner layer diameters not the outer layer diameters, indicating that only in small diameter CNTs are excitons strongly pronounced, and for those the excitons the frequencies will be in the 500-1000 meV regime. This analysis, combined with the examples highlighted in Extended Data Figure 5 and 6, suggests that excitons do not impede the vibrational measurements performed in this manuscript.

However, it is also important to note that the relative chiralities between CNT shells can have significant effects on the splitting and frequency of the excitonic transitions. Comparisons of the optical transition energies to the exact chiralities of the inner and outer walls of a CNT have revealed the inter-wall interactions can change the peak frequencies by 20-30 meV in either direction in most cases, and up to 200 meV in the most extreme cases[17]. Thus, without clear diffraction data to determine the exact chirality of all CNT walls present in the measured tubes, the exact expected frequency for the $S_{11}$ and $S_{22}$ transitions has some potential error. To absolutely exclude the possibility of the infrared peaks attributed to water as non-excitonic clear diffraction patterns from each CNT is necessary.

*Vibrational Electron Energy Loss Spectroscopy (Validation of the Experimental Results):* The project overall consisted of 11 filled CNT samples across 6 sessions. A total of 61 separate CNT samples were examined as part of this study and eight exhibited a signal we attribute to water, resulting in an approximate 13% filling rate was observed across the samples. The spectra attributed to O-H stretches and eight representative spectra of empty CNTs are shown in Supplementary Information Figure S3. On these samples there were five CNTs showed a non-H-bonded peak at the free O-H stretch (peak at ~455 meV) and three exhibiting the bulk-water-like response at the bonded O-H stretch (peak at ~420 meV). The critical point of this figure is that all five CNTs exhibiting a strong free O-H peak, show no trace of a bonded O-H peak, indicating we have found 5 separate nanotubes exhibiting the non-H-bonded response. We note that this observation does not preclude that confined water could potentially have different internal configurations other than the majority non-H-bonded behavior described in Main Text Figure 2, as line shapes and peak frequencies change slightly between the observations.

To achieve statistics many CNTs must be examined, below is a brief summary of the time constraints of the vEELS experiment in a single working day. A standard sample prepared according to the procedure above, generally contains between 5-15 CNTs that can be identified in TEM-mode or highly defocused Ronchigram by searching for charging effects on the SiN membrane near the CNTs. After identifying all CNTs, vEELS is conducted in a STEM mode which can help identify whether the CNTs are isolated single CNTs or bundles, and aloof point spectra can be acquired to determine whether or not the O-H signal is present. Nominally, all CNTs can be identified and have the initial characterization performed in a matter of a few hours, leaving time for in depth characterization (including cryo) of 2-3 CNTs of interest in a standard working day.

We have also performed other experiments not shown in the main text, that help validate the results of the main text analysis. Two key examples are presented here. In Extended Data Figure 5, we show the sole observed occurrence of partial filling. For every other CNT, the same O-H stretch frequency could be observed across the entire CNT, distances of tens of microns, while in this case the water signature was observed strongest by the window edge, and then again far away from the window edge (towards the center of the grid). In the area highlighted as P1 in Extended Data Figure 5a, the O-H stretch peak vanished.

We additionally highlight an occurrence where the CNT burst during acquisition. This is shown in Extended Data Figure 6, which shows a before and after with vEEL spectra from a CNT that broke. It can clearly be seen that the area surrounding the break is highly modified by the contents of the CNT adsorbing onto the surfaces. We note that the material on the surface is likely notice, but some mixture of the water and hydrocarbon contamination from the SiN grid, as it can be seen that the 365 meV C-H stretch emerges strongly after the break and that the spectrum is modified heavily. Moreover, we can see that in the area around the break, the CNT loses the O-H stretch mode afterwards, suggesting that the water escaped from the CNT.

Attempts to repeat the reversibility highlighted in Extended Data Figure 3, were also performed. Given the low filling rate of the CNTs, and a natural variance of the inner diameter of the tubes, finding filled tubes with similar diameters is challenging, and the exact reversible response was not observed again. Nevertheless, we report our observation as an initial experiment demonstrating the potential of vEELS (and especially cryo vEELS) and expect future development of new platforms for confined fluids will help obtain more robust analyses. The analysis is shown in Supplementary Fig. S3-4. Additionally, shown is an observation about the repeatability of the CNT vibrational response across the length of the CNT. Supplementary Fig. S4c shows a spectrum from the 2.3 nm-diameter CNT at 100 K acquired very far away from the original spot, which exhibits both the 400 meV crystalline ice peak and the 420 meV bonded O-H peak as shown in Main Text Figure 3, demonstrating that the response is uniform across the entire CNT. However, as we note in the bottom spectrum, a slight 400 meV peak is also observable very close to the initial position at room temperature (before the initial cooling to 100K). The peak is weaker in the room temperature spectrum, with respect to the 100 K spectrum, and peaks at 400 meV have been observed room temperature water through vibrational sum frequency spectroscopy[18]. As a result, we believe this result still is most consistent with reversible phase change shown for the 2.3 nm-diameter CNT does corresponding to crystalline ice formation. However, this observation highlights the complex nature of the measured vibrational response in vEELS.

*Theoretical Analysis:* The O–H stretching mode was modeled as a one-dimensional quantum oscillator to interpret the vibrational response[19]. The vibrational Hamiltonian is

$$\hat{H} = -\frac{\hbar^2}{2\mu}\frac{d^2}{dr^2} + V(r), \qquad (2)$$

where $\mu = m_{\mathrm{O}}m_{\mathrm{H}}/(m_{\mathrm{O}} + m_{\mathrm{H}}) = 0.948\, m_{\mathrm{u}}$ is the reduced mass of the O–H bond and $V(r)$ is the potential energy curve[19,20].

Potential Energy Model. For the free O–H stretch, a single Morse potential was used,

$$V_{\mathrm{free}}(r) = D_{\mathrm{OH}}\left[\left(1 - e^{-a_{\mathrm{OH}}(r-r_{\mathrm{OH}})}\right)^2 - 1\right], \qquad (3)$$

where $D_{\mathrm{OH}}$ is the bond dissociation energy, $a_{\mathrm{OH}}$ the potential width, and $r_{\mathrm{OH}}$ the equilibrium bond length. Hydrogen bonding was represented by adding a second Morse term centered at the hydrogen-bond acceptor, a distance $R$ from the donor oxygen, forming a perturbed double-well potential[21]:

$$V_{\mathrm{double}}(r) = D_{\mathrm{OH}}\left[\left(1 - e^{-a_{\mathrm{OH}}(r-r_{\mathrm{OH}})}\right)^2 - 1\right] + D_{\mathrm{HO}}\left[\left(1 - e^{-a_{\mathrm{HO}}(R-r-r_{\mathrm{HO}})}\right)^2 - 1\right]. \qquad (4)$$

All potentials were zeroed at their minimum. Parameters were

$$D_{\mathrm{OH}} = D_{\mathrm{HO}} = 8.0 \times 10^{-19}\,\mathrm{J}, \quad a_{\mathrm{OH}} = a_{\mathrm{HO}} = 2.25\ \text{Å}^{-1}, \quad r_{\mathrm{OH}} = r_{\mathrm{HO}} = 0.96\ \text{Å}, \quad R = 3.48\ \text{Å},$$

reproducing the vapor-phase O–H frequency ($\sim 3700\ \mathrm{cm}^{-1}$) while enabling H-bond-induced redshift[20-22].

With the potential defined, the Hamiltonian was diagonalized numerically using a finite-difference discretization, and transition energies were obtained (for double well potential we focus on the left well). The dipole moments $\mu_{j,n\to n+1}$ between adjacent levels were calculated with linear operator, effective charge of $0.5e^-$. Thermal occupation probabilities $P_n(T)$ were applied following the Boltzmann distribution at $T = 300$ K.

Spectra Construction. The complex dielectric function was then obtained as[22-24]

$$\varepsilon(\omega) = \varepsilon_\infty + \sum_j f_j \sum_n \frac{s_{j,n\to n+1}}{\left(\omega_{j,n\to n+1}\right)^2 - \omega^2 - i\gamma_{j,n}\omega} \qquad (5)$$

$$s_{j,n\to n+1} = \frac{\left|\mu_{j,n\to n+1}\right|^2 * \omega_{j,n\to n+1} * 2\mathrm{N} * P_n}{\hbar\varepsilon_0} \quad (6)$$

where $j$ indexes distinct vibrational environments (free or hydrogen-bonded O–H), $n$ the vibrational level within each potential, and $f_j$ the fractional population of each environment. The high-frequency dielectric constant was set as $\varepsilon_\infty = 2$ for water. $\varepsilon_0$ denotes vacuum permittivity. O-H bond number density $N = 6.6 \times 10^{28} m^{-3}$.

To capture damping effects phenomenologically (Fig. 2), a dimensionless parameter $\Gamma = \gamma/\gamma_0$ was used, where $\gamma_0 = 2 \times 10^{13}\ \mathrm{s}^{-1}$ denotes the characteristic inverse lifetime of the O-H stretch[25,26]. Simulations were performed for $\Gamma = 1, 2,$ and $4$, applied uniformly for all environments to illustrate the progressive broadening.

The resulting energy-loss function (ELF), directly corresponding to the experimental vEELS signal, is[17]

$$\mathrm{ELF}(\omega) = \mathrm{Im}[-\varepsilon^{-1}(\omega)]. \quad (7)$$

All calculations were implemented in Python.

The schematics in Fig. 2d were generated by PyMOL software.

**Methods References**

**Acknowledgments**

This research was primarily supported as part of the Center for Enhanced Nanofluidic Transport (CENT), an Energy Frontier Research Center funded by the U.S. Department of Energy (DOE), Office of Science, Basic Energy Sciences (BES), under Award # DE-SC0019112. Vibrational

EELS research was supported by the Center for Nanophase Materials Sciences (CNMS), which is a DOE Office of Science User Facility using instrumentation within ORNL's Materials Characterization Core provided by UT-Battelle, LLC, under Contract No. DE-AC05- 00OR22725 with the DOE and sponsored by the Laboratory Directed Research and Development Program of Oak Ridge National Laboratory, managed by UT-Battelle, LLC, for the U.S. Department of Energy. The authors would also like to thank Dr. Narayana Aluru of the University of Texas, Austin for discussions on modeling and simulations. Our thanks to Dr. Daichi Kozawa from the National Institute for Materials Science in Japan for helpful discussions surrounding excitonic transitions in the CNTs.

**Author Contribution Statement**

Experiments were conceived of and designed by XX, JCI, AM, and JAH. VEELS data was acquired and analyzed by XX, HAW, and JAH with guidance from JM and AM. Samples were synthesized and prepared by XX, MK, YMT, CLR, and MSS. Morse oscillator modeling was conducted by XX with the guidance from AM. MD simulations were conducted by HAW, XJ with guidance from DLB and STP. Manuscript and supplementary materials were written by XX, AM, and JAH. All authors contributed to the editing and revision of the manuscript.

**Author Information Statement**

Correspondence and requests for materials should be addressed to arunava@stanford.edu, or hachtelja@ornl.gov.

**Competing Interests Statement**

The authors declare no competing interests.

**Data Availability Statement**

The raw data and Jupyter Notebooks used to process and visualize the figures in the main text have been made available at the following repository. All other codes and data are available upon request from the corresponding authors.

Extended Data for

# Hydrogen bonding in water under extreme confinement

## Supplementary Figures

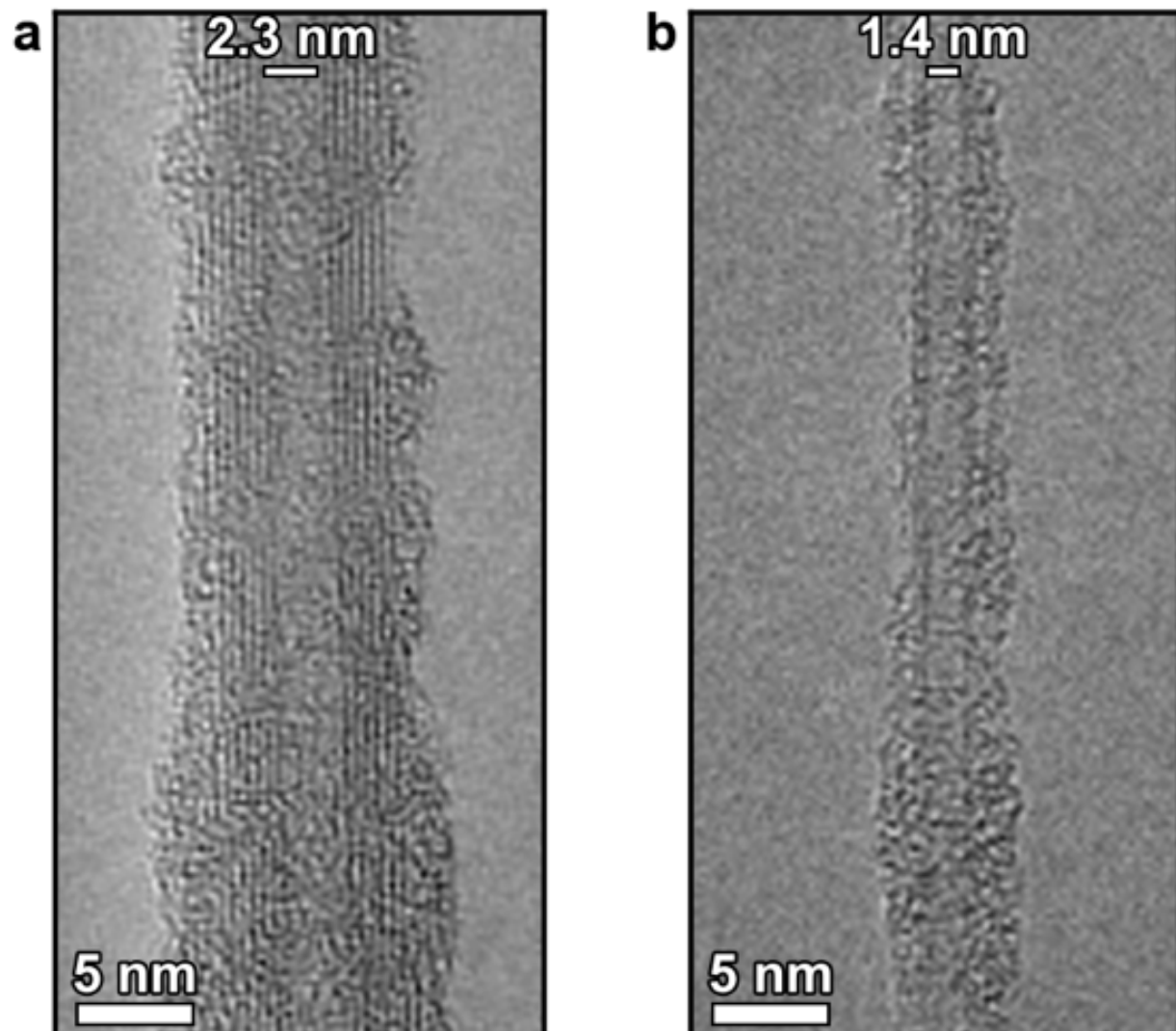

**Extended Data Figure 1. TEM Images of Main Text CNTs.** TEM images of the most examined CNTs in the main text with large and medium diameters. (a) The filled large-diameter tube from Fig. 1d-h, Fig. 2a, and Fig. 3a is measured to have an inner diameter of ~2.3 nm. (b) The filled medium-diameter tube from Fig. 2a and 3a is measured to have an inner diameter of ~1.4 nm.

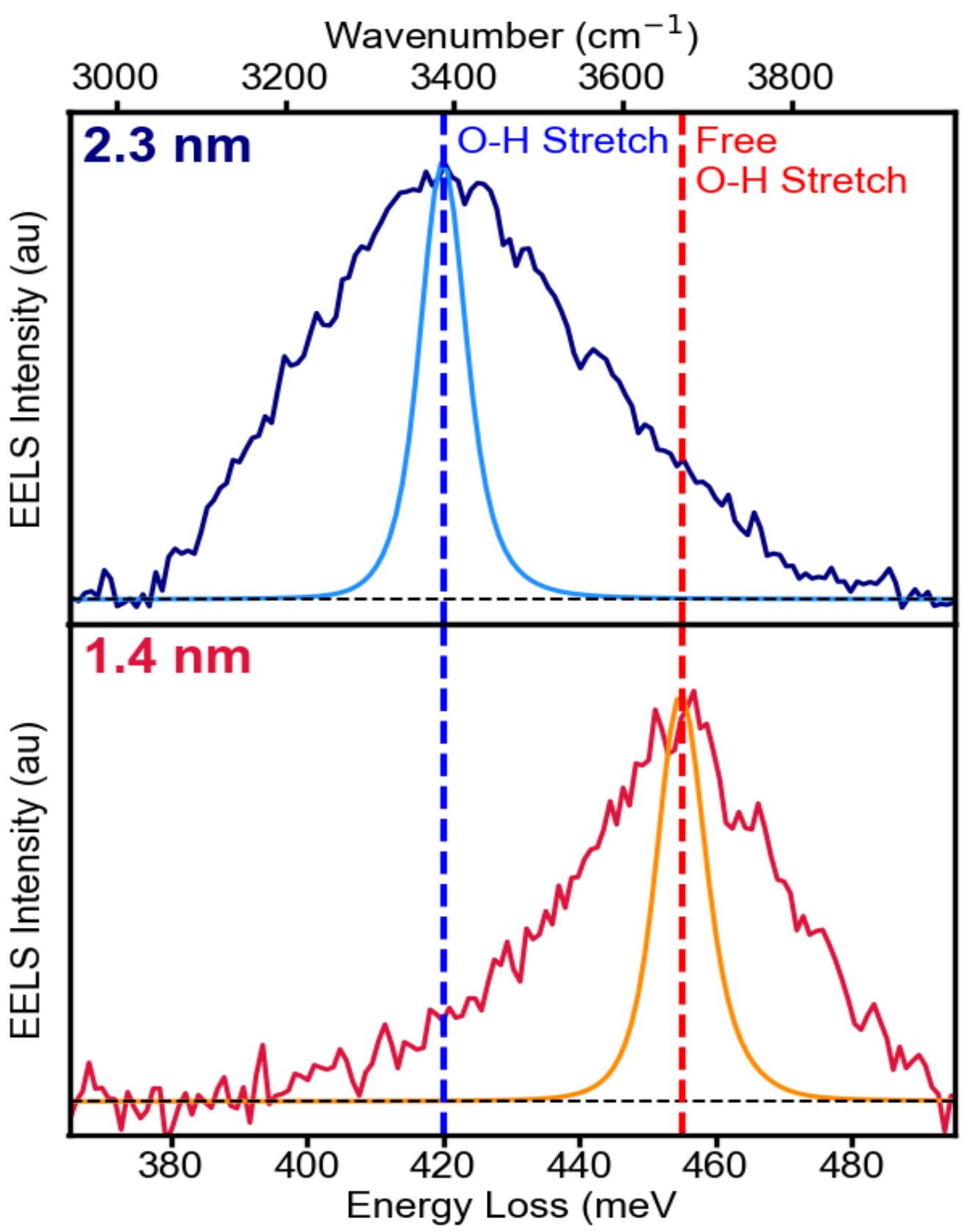

**Extended Data Figure 2. Instrumental broadening vs. Spectral Linewidth.** In this figure we show that our instrumental resolution is significantly below the measured linewidth of the vibrational spectra for the vibrational spectra shown in the large (2.3 nm) diameter tube and the smaller (1.4 nm) diameter tube from the main text figures. For the 2.3 nm tube the full-width at half-maximum (FWHM) of the O-H stretch is 50.1 meV, while the FWHM of the 1.4 nm tube is 34.0 meV. For all vEEL spectra the energy resolution can be quantified by measuring the FWHM of the zero-loss peak (ZLP), which corresponds to all electrons in the EEL that have either elastically scattered off the sample or not interacted with the sample at all. As a result, this peak represents the spread of electron energies in the beam with which all peaks are convolved with and represents the instrumental broadening. For both spectra the FWHM of the ZLP is 8.5 meV, indicating that the vibrational peak width is dominated by the native linewidth of the peak, not instrumental broadening.

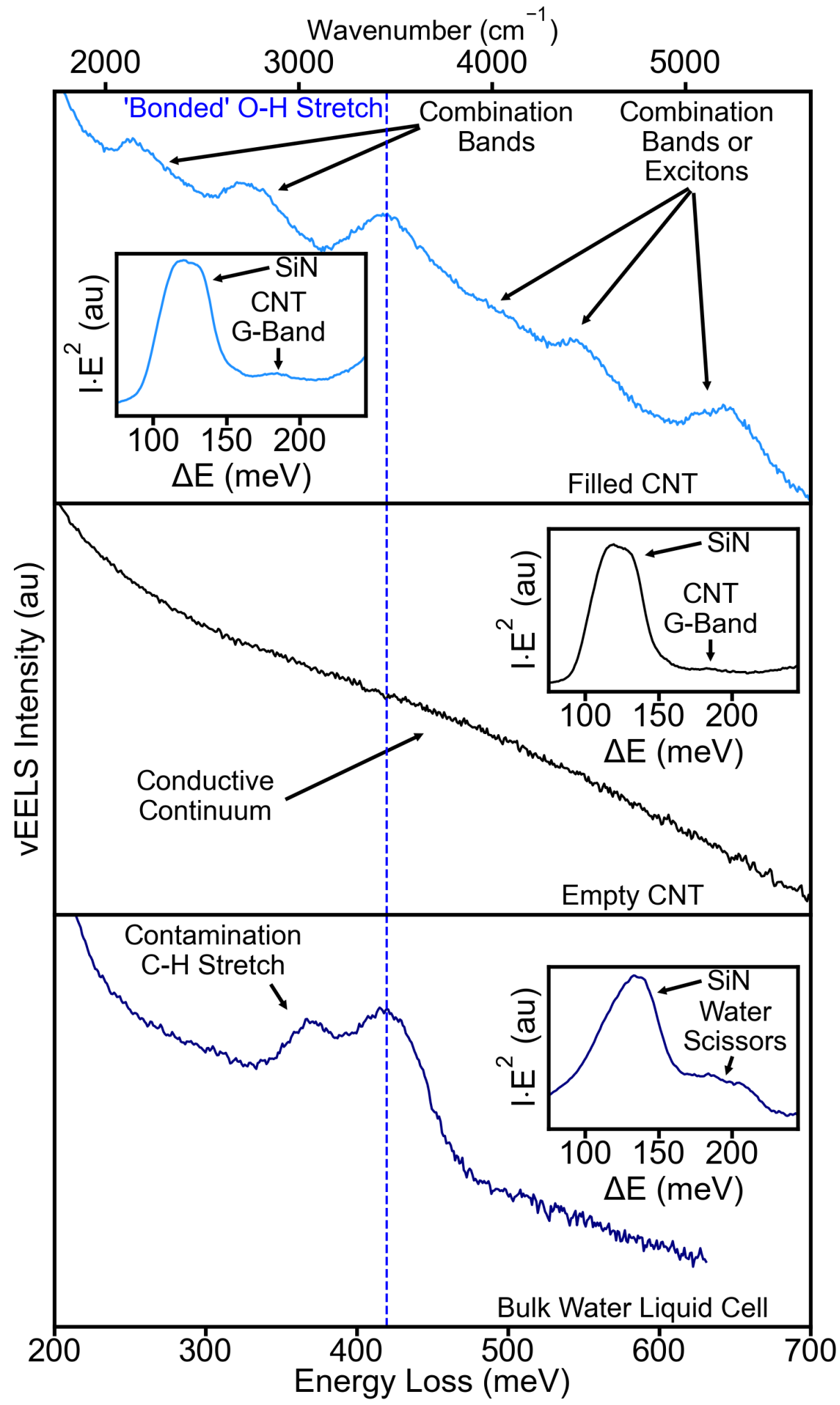

**Extended Data Figure 3. Full vibrational spectrum of Filled CNT, Empty CNT, and Bulk Water.** Here we show the full vibrational response of all three spectra shown in the main text of Figure 1. In each we show the spectrum in log-scale and in the inset we show the SiN/G-Band vibrations of the substrate/CNT. (Top) filled CNT, (Middle) Empty CNT, (Bottom) bulk liquid cell.

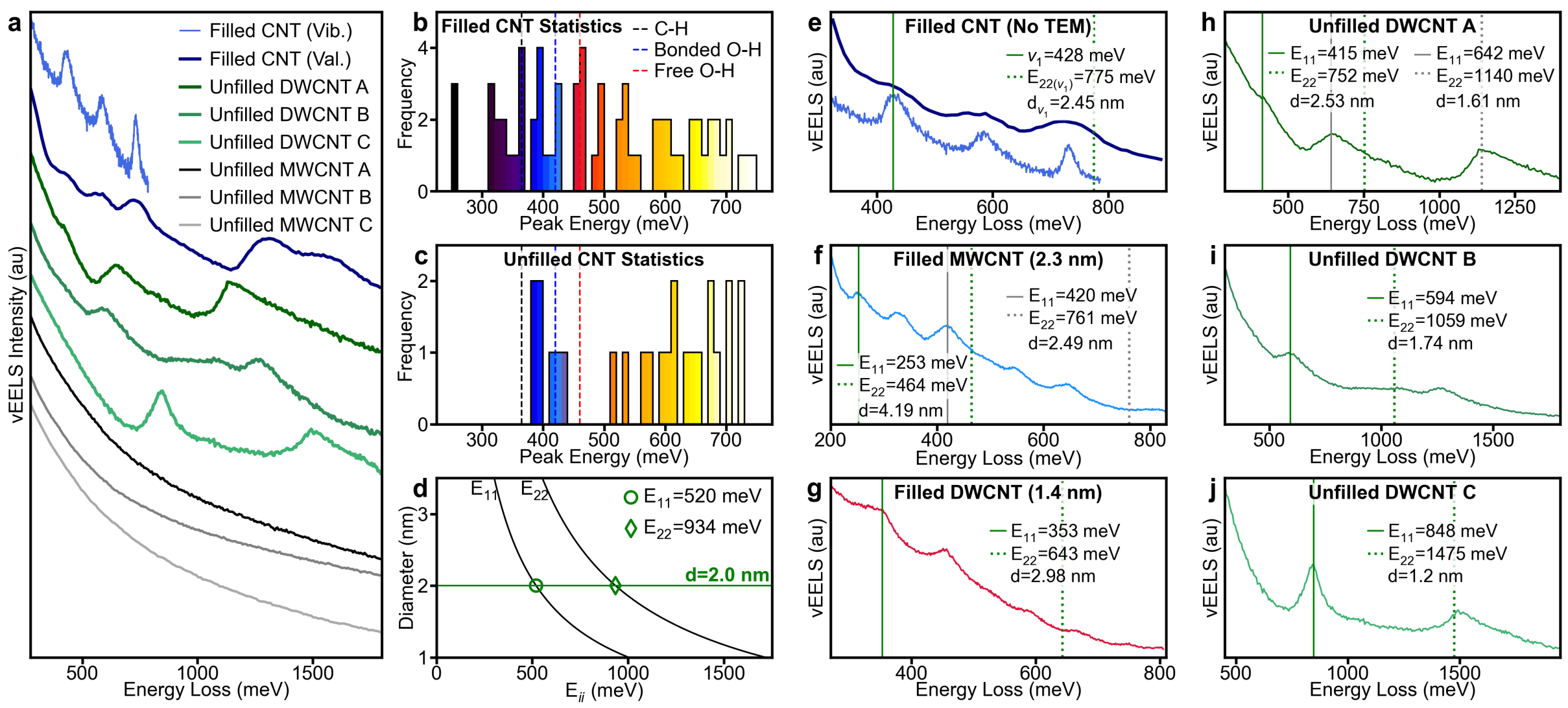

**Extended Data Figure 4. Valence EELS of Filled and Unfilled CNTs.** (a) Vibrational/valence regime EELS for filled CNTs from the confined water sample compared to valence EELS for unfilled commercial CNTs. (b,c) Statistics of the peak distributions for the filled CNTs (b) and the commercial DWCNTs, with estimated inner diameters between 1.3 and 2.0 nm. (c), showing significant difference between excitons and vibrations. (d) Empirical formula for relationship between CNT diameter and excitonic energies derived in Methods Ref. 15. (e,f,g) Vibrational spectra from filled CNTs showing inconsistency with empirical formula. (h,i,j) Valence spectra from unfilled CNTs showing strong consistency with the empirical formula.

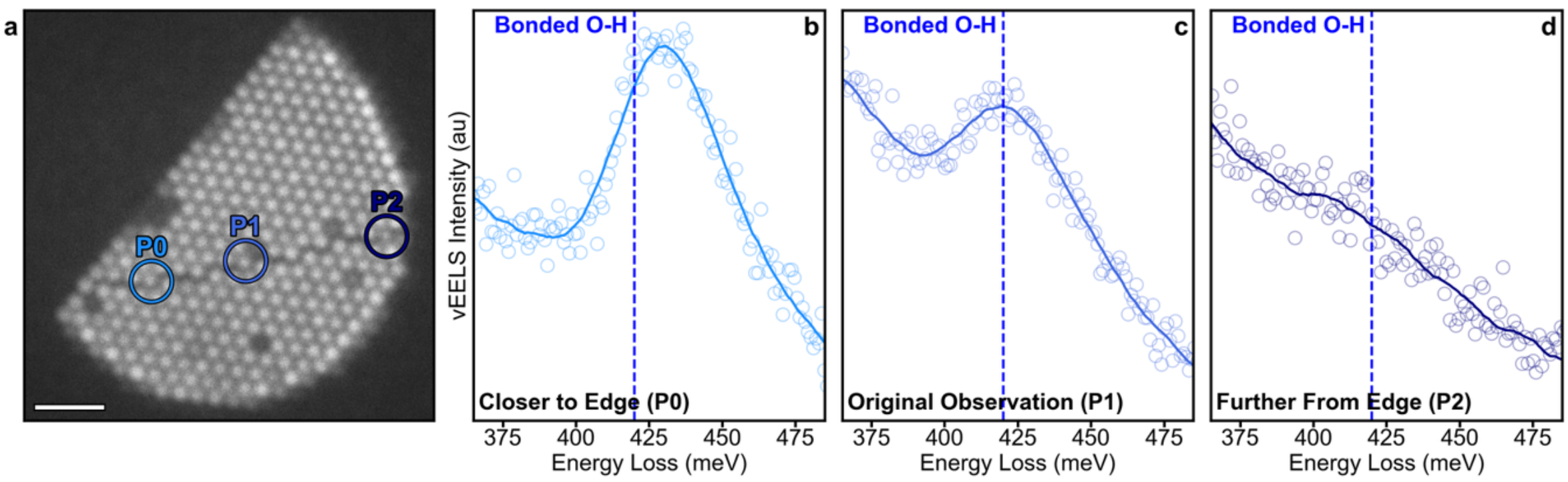

**Extended Data Figure 5. Example of Partial Filling in CNTs.** Here we highlight a measurement of a water-filled CNT where partial filling was observed. (a) Shows a defocused STEM image of the CNT on the SiN holey grid, with three locations highlighted showing a gradual decrease in the observed intensity of the O-H signal (b-d) where additional spectra were measured. The initial measurement and identification of the bulk-like water response is shown in (c) and is consistent with a strong observation of the O-H stretch. Going further away from the CNT wall the bonded O-H stretch was not observed (d) while closer to the edge it was observed with an even higher intensity than in the initial observation(b). While in most CNTs the O-H stretch measurements were consistent across the entire CNT, in this example it was not. This same CNT is shown in other figures as well, namely it is 'bulk-water-like 2' in Fig. S3a, and 'CNT B' in Fig. S4.

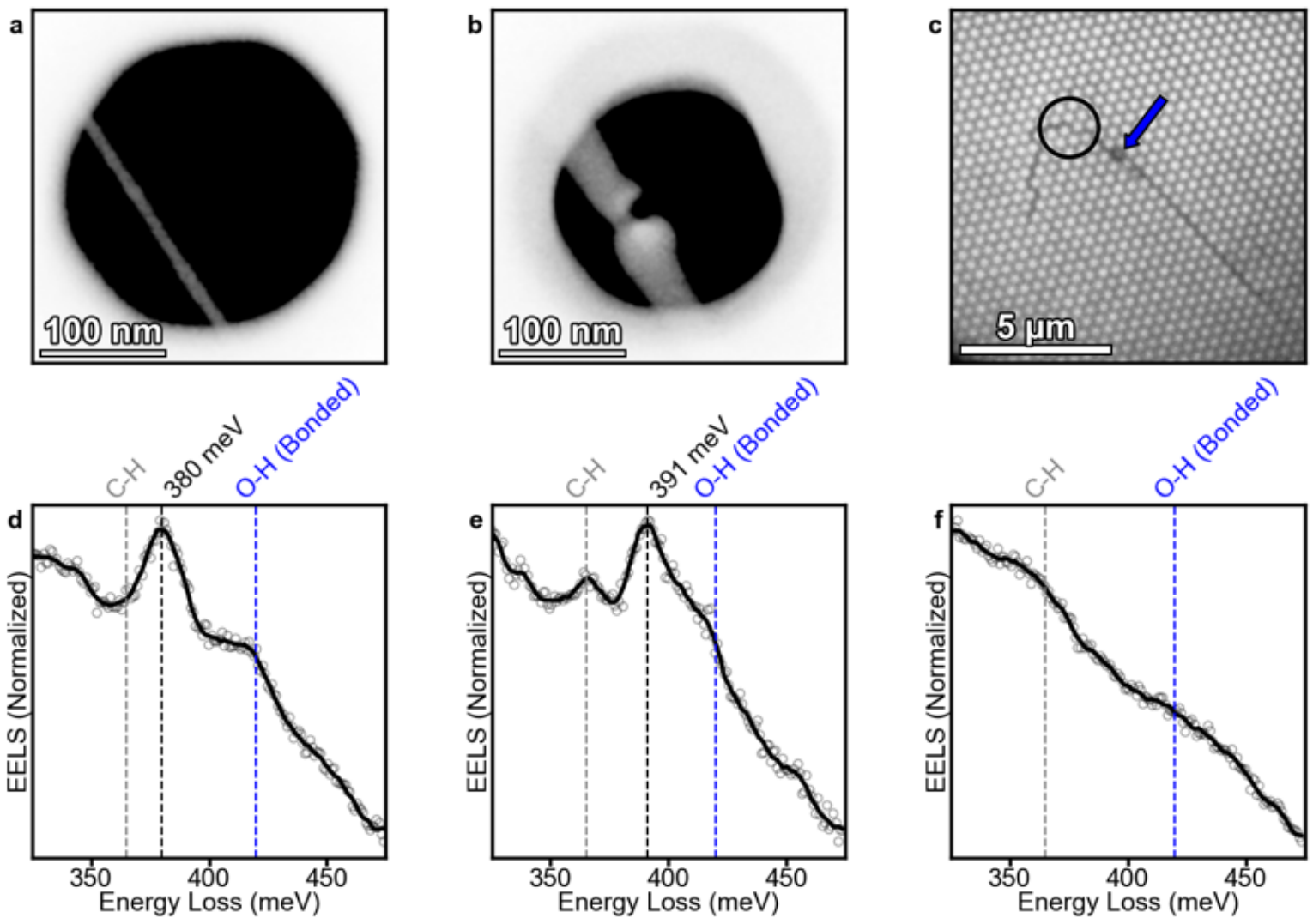

**Extended Data Figure 6. Example of CNT Breaking During Experiments.** Here, we highlight an instance where a CNT broke during the experiment. (a) shows the reference image of the sample before measurement, and (b) shows the reference image after the measurement. Significant buildup of contamination is present in a way that is no present in any of our other measurements. (c) Defocused Ronchigram shows break occurs above position of blowout (blue arrow). (d) CNT spectra from before break. (e) CNT spectra acquired immediately after the break. (f) CNT spectra acquired from area above break (black circle). CNT examined here is bulk-water-like 3 in Fig. S3a.

Supplementary Information for

# Hydrogen bonding in water under extreme confinement

## Supplementary Figures

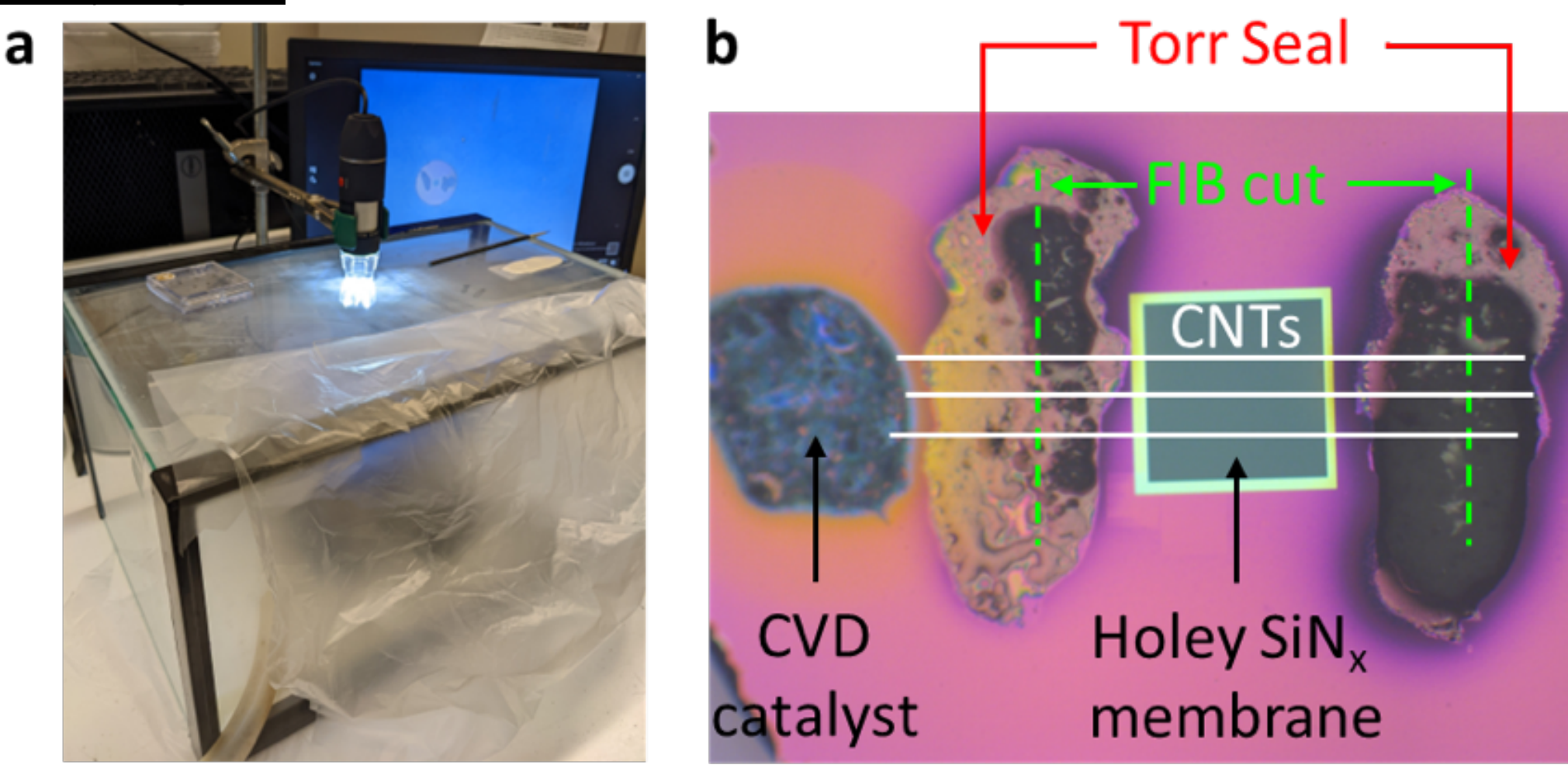

**Supplementary Figure S1. Procedure for CNT water filling**. (a) A custom-built humidity chamber with a microscope. (b) Optical micrograph of a holey $SiN_x$ TEM chip after CVD growth and FIB cutting (along green dashed lines) of CNTs (schematically illustrated as continuous white lines) as well as their exposure to above 99% RH conditions and Torr Seal application. Note that Torr Seal covers the FIB cut regions but not the holey $SiN_x$ membrane.

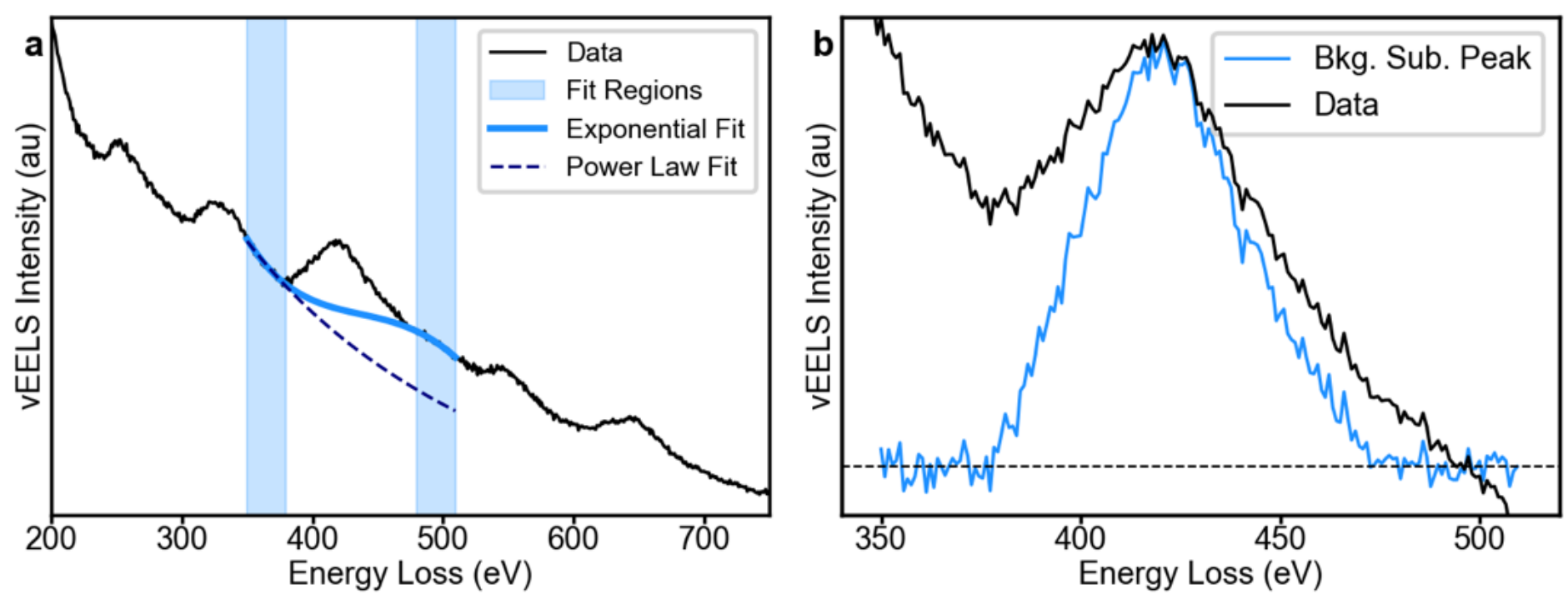

**Figure S2. Background Subtraction in vEELS.** (a) A vEEL spectrum with the fit regions for a two-region, third-order exponential fit, and the resulting background. (b) Comparison of the background subtracted data and the actual peak.

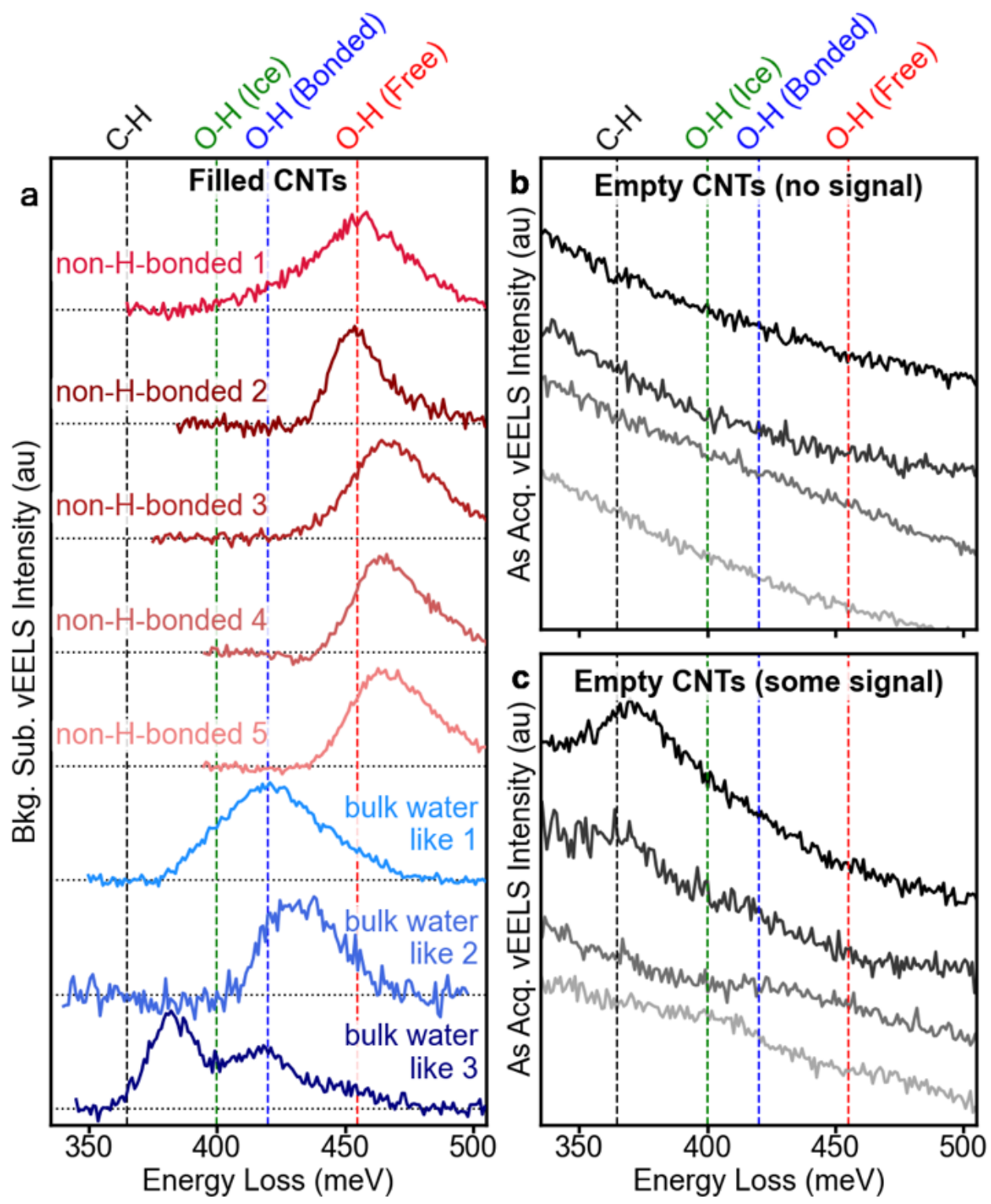

**Figure S3. Statistics of the Water-Filled CNT Response.** Here, we show a representative vibrational spectrum from each of eight filled CNTs that were observed across all experimental sessions (a), along with eight (out of 53 total) empty CNTs to illustrate the observed variance in the observed signal in this regime (b,c). Both the non-H-bonded and the bulk-water-like vibrational responses were observed on three separate samples on three separate sessions. The majority of the empty CNTs show no peaks whatsoever in the 350-500 meV region (31 of 53), some exhibit faint peaks across this range (or even sharp peaks at the C-H stretch from hydrocarbon contamination). All analysis was focused on samples exhibiting definitive signals. *Non-H-bonded 1* and *bulk-water-like 1* are the two spectra highlighted in the main text in Figure 2/3 and were observed on the same sample during the same session where cryogenic measurements were performed. The *non-H-bonded 3-5* spectra were acquired during the very first session, all from the same sample, however this sample was a mesoporous Si substrate. On this grid, the tubes were frequently bundled together, and the random variations of the membrane made it impossible to track down the same CNTs later for TEM analysis. All other experiments were conducted on periodic SiNx hole grids which enabled the CNTs to be found later for TEM.

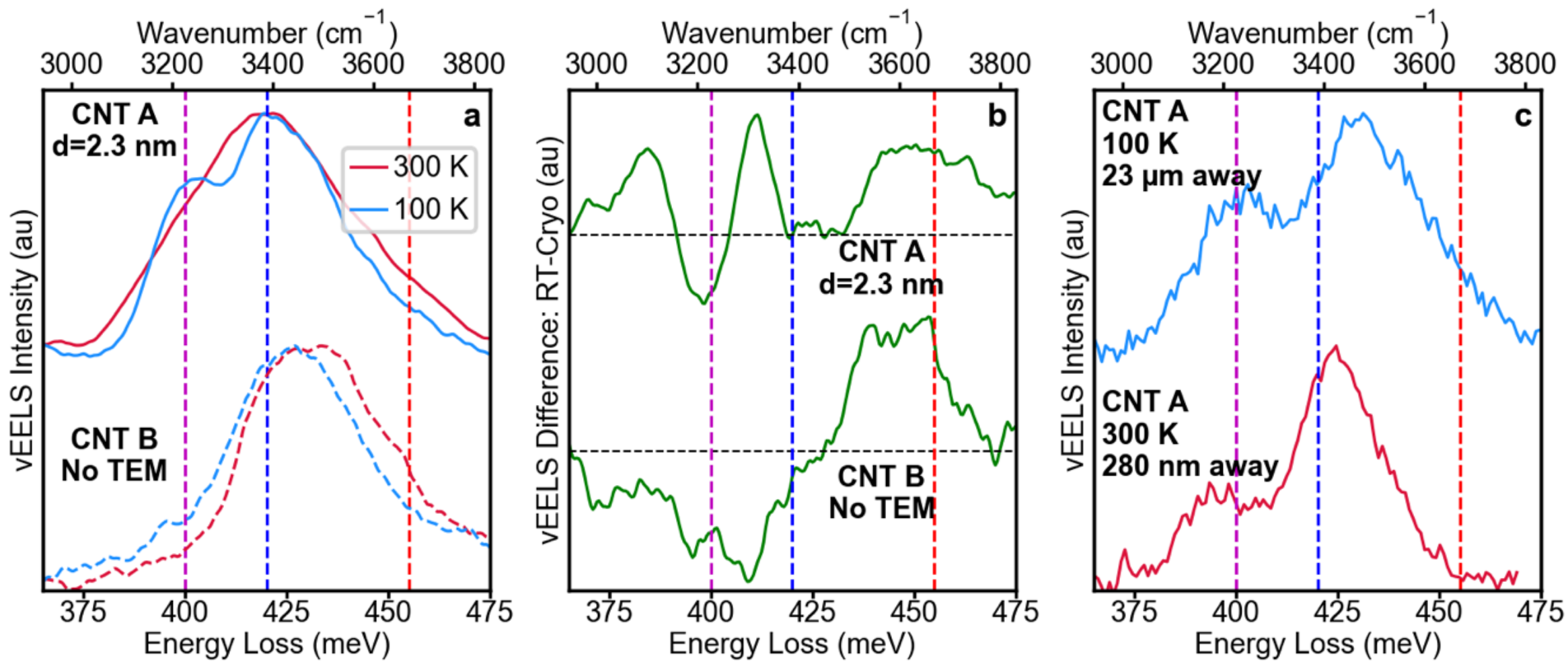

**Supplementary Figure S4. Repeatability of phase change signatures in large diameter CNTs.** (a) Comparison of the room temperature (300 K) and cryogenic (100 K) spectra for the CNT shown in the main text (CNT A) and another one (CNT B). (b) Difference spectra of the room temperature and 100 K spectra for each CNT. (c) shows additional measurements on the 2.3 nm CNT. The top (blue) spectra shows a cryo measurement conducted many microns away from the original position of the spectra and the bottom (red) spectra shows a room-temperature measurement very close to the original position.

**Molecular Dynamics Simulations**

Given the state of the art of molecular-dynamics (MD) simulations of the vibrational spectra of water confined in CNTs[1-3], we performed MD simulations using density-functional-theory (DFT)-based machine learning interatomic potentials (MLIPs; see Methods section below) to systematically investigate the relationship between confinement geometry and water vibrational behavior. Figure S5 presents vibrational density of states (vDOS) spectra as a function of three key parameters: nanotube diameter (Fig. S5a), water density (Fig. S5c), and temperature (Fig. S5e). These simulations provide insight into how the characteristic O-H vibrational signatures evolve with changing confinement conditions.

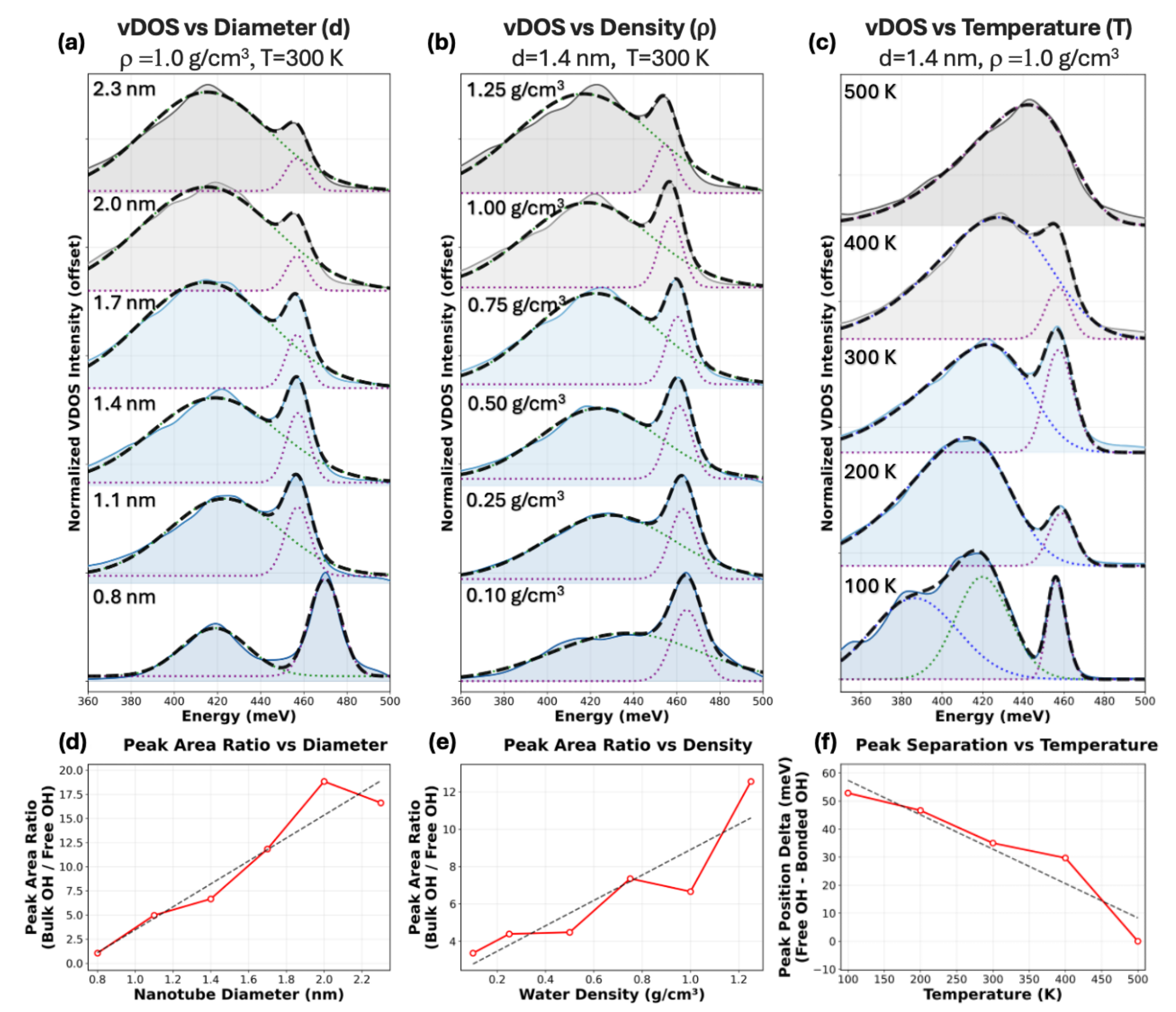

**Supplementary Figure S5**: **VDOS trends of water in carbon nanotubes (MD + Gaussian fits). (a,b,c)** Filled curves from MD simulations are overlaid with black dashed curves representing Gaussian fits. Each fit is decomposed into free OH (purple dotted) and bonded-OH (green dotted), and an extra bonded-OH mode arising from a low-temperature phase transition (blue dotted), possibly to a form of ice. Panel (a) varies nanotube diameter (0.8-2.3 nm), (b) varies water density (0.10-1.25 g/cm$^3$ at 1.4 nm), and (c) varies temperature (100-500K at 1.4 nm). In each case, the other two parameters are as indicated in the top labels. (d,e) Quantification of fit results: the peak-area ratio of bonded to free OH grows nearly linearly with increasing diameter and density), respectively. (f) The separation between free and bonded-OH peaks decreases monotonically as temperature rises, reflecting thermal broadening and increased anharmonicity as temperature increases. Together, these panels reveal how geometric confinement, molecular packing, and thermal effects influence H-bond networks and why real single-CNT vEELS spectra, subject to sample heterogeneity, can be so diverse.

The diameter dependence (Fig. S5a) captures the water-configuration transition from quasi-one-dimensional water chains in narrow CNTs to more bulk-like behavior in wider confinements, where competition between the CNT-water and water-water interactions fundamentally alters both the geometrical arrangement of water molecules and the H-bond network. Fig. S5b explores water-density effects at a fixed diameter of 1.4 nm, revealing how water packing (akin to pressure) within the cylindrical geometry influences the vibrational modes. The temperature series Fig. S5c probes the thermal evolution of these confined structures, providing a controlled comparison to the possible heating effects that may occur in experiments.

Tuning of each of these parameters leads to an exchange of intensity between the bulk O-H and free O-H peaks in the simulated vDOS in a near-linear fashion. While we do not find a single peak as in the experimental spectra, the systematic trends provide valuable insights into the behavior of pristine CNTs and show that small changes in confinement conditions can lead to substantially different spectral signatures. This sensitivity underscores why the structural and chemical heterogeneities present in real CNT samples, as probed by single-CNT vEELS, can produce diverse experimental spectra.

Single-CNT vEELS measurements inherently probe the structural and chemical imperfections of individual CNTs that are averaged out in ensemble studies. Unlike bulk optical or neutron scattering experiments, vEELS directly correlates spectral signatures with specific regions of individual CNTs, which means that local inhomogeneities such as uneven filling, contaminants, defects, and diameter fluctuations may modulate H-bond networks and their vibrational fingerprints. While beam-sample interactions could introduce additional complexity through local heating or radiolysis, we observe no notable differences between aloof and on-sample spectra, indicating minimal beam-induced artifacts. This sensitivity to sample heterogeneities represents both an opportunity and a challenge: single-CNT vEELS provides unprecedented insight into confinement physics but demands rigorous statistical sampling to distinguish intrinsic behavior from sample variability.

*Methods*

To investigate the influence of confinement in CNTs on the vibrational spectra of water, we employed a multi-scale computational approach combining DFT calculations with MLIPs and extensive MD simulations. This workflow enables us to achieve the long timescales necessary for converged vibrational spectra while maintaining near-ab initio accuracy in the description of hydrogen bonding and water-carbon interactions. We first performed DFT-based MD simulations on representative water-CNT configurations to generate a diverse training dataset capturing the potential energy surface across varying confinement geometries, water densities, and temperatures. These DFT data were then used to train a MACE machine learning potential[4], which was subsequently deployed for MD simulations to compute vDOS spectra. This section details the structure models, DFT simulation parameters, MLIP training procedure, and Molecular Dynamics employed in this study.

Water-in-CNT structure models*:* The density of water in a CNT is defined as the ratio of the mass of water molecules to the volume occupied by water in the CNT, estimated by removing a 2-Å-wide shell from the CNT radius defined by the carbon nuclei. To study partial filling (low densities) it was found necessary to use shorter CNTs (2 nm), as the water tends to form droplets of an equilibrium density (~1 g/cm$^3$) when longer nanotubes contain sufficient water. As expected, vDOS of droplets in CNTs give a bimodal vDOS spectrum that depends on the surface-to-volume ratio. All CNTs studied in this work are armchair CNTs with no defects and are assumed periodic along the tube axis. All CNTs are allowed to vibrate during the MD simulations. Nanotubes of length 2nm, diameter 0.8, 1.1, 1.4, 1.7, 2.0, and 2.4 nm were created with water densities 0.1, 0.25, 0.5, 0.75, 1.0, and 1.25 g/cm$^3$.

DFT simulations: We performed DFT-MD simulations for water confined in carbon nanotubes (CNTs) using the Vienna ab initio simulation package (VASP) and the projected augmented wave (PAW) method. The exchange-correlation functional for electrons was described by the generalized gradient approximation (GGA) in the Perdew-Burke-Ernzerhof (PBE) form, with the Gimme D3 method included to describe van der Waals (vdW) interactions. This choice of functional was informed by the review performed by Brandenburg and Michaelides[5]. The simulations were conducted in the NVT ensemble using the Nosé-Hoover thermostat and a 0.5 fs timestep. Simulations were performed to provide configurations for training and validating machine-learned potentials (MLIPs). Select vDOS calculations were also performed to further validate our MLIPs (see section on MLIPS below). Convergence of the water and CNT temperatures to a single equilibrated temperature was confirmed before the calculation of vDOS was initiated.

Machine Learned Interatomic Potential (MLIP): The DFT simulations were processed into a dataset of 22,613 configurations for the training set, 2,261 configurations for the validation set, and 5,668 configurations for the test set. A MACE machine-learning interatomic potential was trained[6]. The MACE architecture employs equivariant message passing with body-ordered atomic cluster expansions up to correlation order 3 (4-body interactions), enabling accurate representation of many-body interactions essential for hydrogen bonding and confined water dynamics. The model architecture consisted of hidden irreducible representations of 128×0e + 128×1o, with a maximum angular momentum channel of L=3, a radial cutoff of 6.0 Å and 2 message passing iterations. Performance of the trained model on the test set is visualized in Figure S6a,b. The model exhibits a RMSE of 0.048 meV/atom for energy predictions and a RMSE of 0.005 eV/Å for force predictions. A DeepMD potential was also trained for comparison using the se_e2_a atomic descriptor[7]. While this model achieved higher inference speeds compared with the MACE model, we ultimately chose accuracy over speed. Though the DeepMD potential was not employed for results reported here, validation is provided for interested readers.

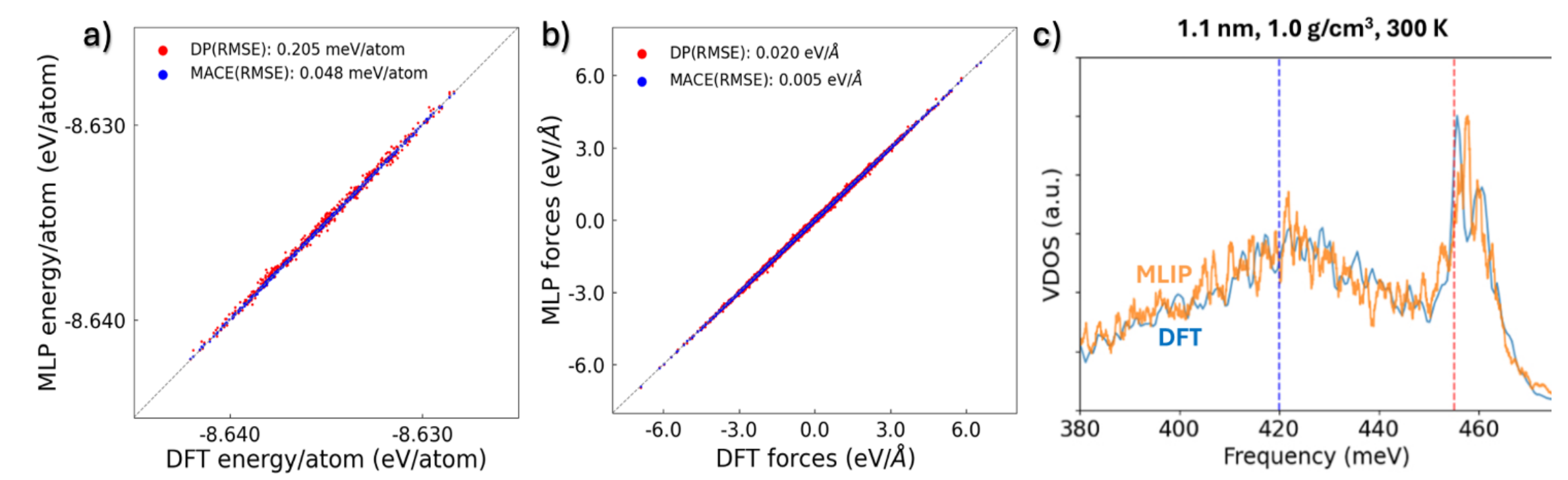

**Supplementary Figure S6**: **Machine learning potential performance.** Parity plots comparing DFT-calculated and ML-predicted (a) energies per atom and (b) atomic forces for both DeepMD and MACE models. MACE achieves an RMSE of 0.48 meV/atom for energies and 0.005 eV/Å for forces, while DeepMD achieves 0.205 meV/atom for energies and 0.026 eV/Å for forces. MACE demonstrates significantly superior force accuracy, which is critical for reliable molecular dynamics trajectories, and was therefore selected for all production simulations. In panel (c) VDOS is obtained from the Fourier transform of the velocity autocorrelation functions computed from DFT-based (blue) and MLIP-based (orange) MD simulations. Under the selected conditions, namely, 1.1 nm CNT diameter, water density of 1.0 g/cm³, and temperature of 300 K, the VDOS profiles from DFT and MLIP exhibit excellent agreement. The blue and red dashed lines denote two characteristic O-H stretching frequencies observed in different environments: ~420 meV for hydrogen-bonded O-H vibrational modes (as in bulk water) and ~455 meV for free O-H stretching modes.

MLIP Molecular Dynamics (MD): MLIP-MD simulations were conducted utilizing the LAMMPS software. Simulations were equilibrated over 100 ps within the NVT ensemble using the Nosé-Hoover thermostat and a 0.5 fs timestep. Once temperature equilibration was achieved system-wide, the simulations were run for an additional 100 ps, during which data was collected. In addition to the traditional static validations, which measure the difference between DFT and MLIP atomic energies and forces, we performed dynamic validation, which primarily focuses on the vDOSv obtained by DFT and MLIP. As shown in Fig. S6c, the DFT and MLIP vDOS for a test case of 1.1 nm, 1.0 g/cm$^3$ inthe frequency range of interest exhibit perfect agreement.